\documentclass{ieeeaccess}
\usepackage{cite}
\usepackage{amsmath,amssymb,amsfonts}
\usepackage{algorithmic}
\usepackage{graphicx}
\usepackage{textcomp}
\usepackage{color, colortbl}
\usepackage{adjustbox}
\usepackage{booktabs}
\usepackage{tabularx}
\usepackage{subcaption}
\usepackage{makecell}

\usepackage{array}
\usepackage{caption}
\usepackage{multirow}
\usepackage{url}
\usepackage{float}
\usepackage{hyperref}
\usepackage{adjustbox}
\usepackage{rotating} 
\usepackage{array}
\usepackage{makecell}
\usepackage{afterpage}
\usepackage{supertabular}
\usepackage{longtable,lscape}
\usepackage{multirow}
\usepackage{enumitem}
\hypersetup{colorlinks, citecolor={black}, filecolor=black, linkcolor=black, urlcolor=black}

\definecolor{red}{rgb}{0, 0, 0}

\def\BibTeX{{\rm B\kern-.05em{\sc i\kern-.025em b}\kern-.08em
    T\kern-.1667em\lower.7ex\hbox{E}\kern-.125emX}}
\begin{document}
\history{Date of publication xxxx 00, 0000, date of current version xxxx 00, 0000.}
\doi{10.1109/ACCESS.2020.DOI}

\title{Smartphone Sensing for the Well-being of Young Adults: A Review}

\author{\uppercase{Lakmal Meegahapola}\authorrefmark{1,2},
\uppercase{Daniel Gatica-Perez\authorrefmark{1,2}}}
\address[1]{Idiap Research Institute, Martigny, Switzerland}
\address[2]{\'Ecole Polytechnique F\'ed\'erale de Lausanne (EPFL), Lausanne, Switzerland}
\tfootnote{This work was funded by the European Commission’s Horizon 2020 project "WeNet: The Internet of Us", under grant agreement 823783.}

\markboth
{Meegahapola and Gatica-Perez}
{Smartphone Sensing for the Well-being of Young Adults: A Review}

\corresp{Corresponding author: Lakmal Meegahapola (e-mail: lmeegahapola@idiap.ch)}

\begin{abstract}
Over the years, mobile phones have become versatile devices with a multitude of capabilities due to the plethora of embedded sensors that enable them to capture rich data unobtrusively. In a world where people are more conscious regarding their health and well-being, the pervasiveness of smartphones has enabled researchers to build apps that assist people to live healthier lifestyles, and to diagnose and monitor various health conditions. Motivated by the high smartphone coverage among young adults and the unique issues they face, in this review paper, we focus on studies that have used smartphone sensing for the well-being of young adults. We analyze existing work in the domain from two perspectives, namely Data Perspective and System Perspective. For both these perspectives, we propose taxonomies motivated from human science literature, which enable to identify important study areas. Furthermore, we emphasize the importance of diversity-awareness in smartphone sensing, and provide insights and future directions for researchers in ubiquitous and mobile computing, and especially to new researchers who want to understand the basics of smartphone sensing research targeting the well-being of young adults. 
\end{abstract}


\titlepgskip=-15pt

\maketitle

\section{Introduction}\label{sec:introduction}
Smartphones have been rapidly evolving during the past decade due to advancement of technology in a multitude of disciplines such as hardware (CPU, GPU) \cite{Deng2019}, sensing \cite{Meegahapola2019a, Jayarajah2018}, computer vision \cite{Howard2017, Vidanapathirana2019}, \cite{Rathnayake2020}, deep learning \cite{Wang2018, Meegahapola2019c}, and human-computer interaction \cite{Raento2009, Coles-Kemp2018, Meegahapola2017}. These advancements, together with the benefits they provide have made smartphones integral components of the lives of people. A study shows that smartphone coverage in young adults aged 18-29 in USA is 96\% \cite{PewResearchCentre2019}, and other analyses provide insights as to how smartphones are affecting human behavior while elucidating the close connection of smartphones to millions of people \cite{BankMyCell2019}. Smartphones are more user friendly and interactive, while also capable of collecting and processing contextual data in real-time \cite{Digitaltrends2019}. In addition, app distribution platforms such as Google Play Store \cite{GooglePlayStore2019} and Apple AppStore \cite{AppleAppStore2019} have enabled application developers and researchers to distribute smartphone apps to millions of people worldwide.

Wearable sensing in health monitoring emerged as a trending topic two decades ago with researchers focusing on using wearable sensors for monitoring behavioral patterns, health conditions, and lifestyles \cite{Farringdon1999,Park1999,Pentland2000}. The use of that research in real world settings was rare due to many reasons such as the high cost involved in creating wearable devices, the mindset of people regarding wearable devices, and the inability to distribute devices to wide populations. Hence, many of the research efforts were done in controlled lab settings. With the widespread use of mobile phones in the 2000s, several of these issues went out of the equation as more people, specifically \emph{young adults}, embraced mobile phones. This spanned an emerging literature regarding the utility of mobile phone sensing for large scale applications. A pioneering study on this regard is Reality Mining \cite{Eagle2006} that demonstrated the utility of phone sensing to collect contextual data passively (gps, bluetooth traces, app usage, charging events), in the wild (out of the lab setting, with 100+ people), for an extended period of time (1 year). Works such as UbiFit Garden \cite{Consolvo2008} and MyExperience \cite{Froehlich2007} further demonstrated the capabilities of mobile phones in processing data obtained with external and internal sensors combined with self-reports for behavioral analysis. The emergence of smartphones with more sensing capabilities and interactions compared to traditional phones injected new momentum into mobile sensing research with a shift towards \emph{Smartphone Sensing} \cite{Lane2010}.

\subsection{The Well-being of Young Adults}
People of various age groups have different lifestyles, behavioral patterns, thought processes, and biological characteristics \cite{Rizzutoe2012,Cohen2012}. Young adults (even though there is not a unique definition for the age of young adults, we follow the age range 18-35 suggested by Petry et al. \cite{Petry2002}) go through different circumstances in life compared to older generations, and this is reflected in the activities, social interactions, dietary habits, and even the physical and mental health conditions they have to face \cite{Patel2007,Houstan2001,Newcomb1988,Tabuchi2013,Sung2019,Nadai2019}. Many young adults are doing their studies, in their early career, in the first few years of marriage, unemployed, or a combination of the aforementioned. Considering this stage in life, it is known that stress, anxiety, depression, obesity, alcohol/smoking/drug addictions, and unhealthy food habits are common among young adults \cite{Sung2019} \cite{HealthLinkBC2019,HealthLinkBC2019v2,Pelletier2014}, and the reasons why they face these issues might be different to why someone from another age group would face the same type of issue. For example, it is common to see undergraduate students having depression/stress due to steep work load they have in university, while for a person aged 40-50, they might have depression/stress due to family or job related conditions. Further, young adults use social media and smartphones far more than older generations \cite{PewFactSheet2019}, and prior research suggests differences on the way people use the phone depending on age \cite{Andone2016, Alexander2015}. In addition, even international organizations such as UNICEF have emphasized the need for human-centered design of products and services for young people \cite{Unicef2020}. Hence, the criteria to quantify various health/well-being conditions of people using smartphone sensing would in principle be different, and needs special consideration for a range of smartphone-related issues starting from sensor selection, app design, and deployment strategies to data analysis techniques, while keeping in mind the \emph{diversity} existing within each age group. Therefore, in this paper, we focus on smartphone sensing research that has dealt with health and well-being, specifically of \emph{young adults}. We detail the state-of-the-art in smartphone sensing studies and discuss the strengths and weaknesses of approaches. Further, we propose a taxonomy for smartphone sensing studies that reflects changes in smartphone usage behaviors to highlight areas in which young adults can be focused. Moreover, we emphasize the need for diversity-awareness in building health-related sensing paradigms, identify research gaps, and discuss future directions of research to advance the field.

\subsection{Background and Motivation}\label{subsec:motivation}

\textbf{Previous Surveys on Smartphone Sensing.} After Chen and Kotz's early survey \cite{Chen2000} regarding mobile sensing, one of the best known reviews in smartphone sensing is by Lane et al. \cite{Lane2010} where they provided details regarding sensors of the smartphone that can be utilized for sensing, application eco-systems, and sensing paradigms. McCracken et al. \cite{McCracken2016} studied the use of optical smartphone sensing in resource-limited settings, reviewing optical sensors available in smartphones and highlighting important aspects regarding the limitations of this domain. In another survey, Christin et al. \cite{Christin2011} analyzed privacy aspects of smartphone sensing in existing applications and identified some of the limitations. Some other reviews in the recent past that involve smartphone sensing discuss domains such as internet of things \cite{Kamilaris2016}, incentive mechanisms for smartphone based crowd sensing \cite{Zhang2016}, dynamics of mobile cloud computing \cite{Fernando2013}, usage of smartphone and mobile sensing data for urban studies \cite{Calabrese2014}, monitoring human movement \cite{Rosario2015}, and smartphone sensing in different application domains \cite{Khan2013}. 

\noindent \textbf{Previous Surveys on Smartphone Sensing for Well-being.} Triantafyllidis et al. \cite{Triantafyllidis2017} provided a comprehensive analysis regarding smartphone usage in healthcare where they emphasized the trend of pervasive healthcare (health anywhere, anytime using smartphones) instead of the current institutional based model which is a burden for older generations due to the high prevalence of chronic diseases. Because the review is about general health/well-being, studies on older generations, infants, and young adults are discussed, and these studies use sensing modalities in different ways to capture different behaviors of these user groups. Cornet et al. \cite{Cornet2018} conducted a similar survey regarding smartphone based passive sensing modalities for general health and well-being. Going a step further from \cite{Triantafyllidis2017}, the review in \cite{Cornet2018} categorized the well-being literature based on different types such as mental health, sleep, general health/well-being, and other domains. This analysis does not include details regarding how sensing modalities can be used in tandem with other self-reported data, and lacks in-depth analysis regarding the use of sensing modalities. Harari et al. \cite{Harari2017} discussed usage of smartphone sensing for studying behavior of everyday life in a brief review. They have categorized several sensing modalities based on the type of sensed behavior (e.g.: physical movements, social interactions, daily activities). Aung et al. \cite{Aung2017} surveyed the literature regarding sensing mental health related symptoms and intervention techniques. Trifan et al. \cite{Trifan2019} described smartphone sensor types used in different health and well-being related conditions in their survey. Further, Mohr et al. \cite{Mohr2017} surveyed machine learning and feature transformation techniques used in mobile sensing research targeting mental health. The focus of this survey was wider because it considered all types of ubiquitous sensors. Finally, Can et al. \cite{Can2019} surveyed literature about detecting stress using smartphone and wearable sensing techniques.

As discussed above, no review/survey paper on smartphone sensing and mobile sensing discusses studies focusing on the diversity attributes of people (e.g.: age, gender, country, etc.). In our view, this is an important domain for researchers, as diversity-aware smartphone sensing paradigms are essential to design comprehensive studies about the behavior of different user groups, conduct more in-depth analyses, and leverage human diversity to create rich, human-centered applications. With smartphones becoming more sophisticated and feature rich, the group of people who have benefited and been affected the most are undoubtedly young adults, while smartphone coverage among young adults is an ever increasing figure \cite{PewResearchCentre2019}. In our opinion, leveraging this smartphone ubiquity among young adults by careful \emph{data} acquisition and \emph{system} design considerations would enable to conduct studies to deepen our understanding of young adults's practices and needs, and the development of better informed systems for support and interaction.

\subsection{Scope of the paper}\label{subsec:scopeofpaper}

The following inclusion criteria (IC) were chosen in order to define the scope of the paper.

\begin{itemize}[wide, labelwidth=!, labelindent=0pt]

\item[\textbf{IC1: Passive Sensing --}]The smartphone sensing studies in discussion should have used at least one passive sensing modality. Smartphone based studies that only utilize diaries, recommendation engines, or questionnaires without any passive sensing were discarded in order to focus on the sensing capabilities of modern smartphones.

\item[\textbf{IC2: Using other data sources --}]Studies might have used external sensing modalities (e.g.: smartwatch, activity trackers, wrist bands, etc.) or external data sources (e.g.: foursquare, instagram, weather data, etc.). However, the main theme of the study should be based on a smartphone sensing app, and external data sources serve as complementary data to enhance the analysis or to improve performance.

\item[\textbf{IC3: Venue of publication --}]We considered publications in computer science, behavioral science, medicine, and psychology venues. While analyzing computer science applications from the contribution they make to the domain, we draw findings from literature of other domains to understand how these findings can be adapted towards improving ubiquitous and mobile computing research.

\item[\textbf{IC4: Targeting Young Adults --}]Young adults are the main beneficiary of the study, or at least have been part of the study cohort. We included studies that have not particularly mentioned a target audience, but used a young adult population in the cohort. This criteria also narrows the scope of the review in terms of the type of health and well-being issues addressed (e.g. not including aspects such as heart diseases, Alzheimer's, stroke, etc.). 

\item[\textbf{IC5: Datasets --}]Some smartphone sensing datasets have been used in several studies over the course of 5 - 6 years. In such cases, some studies that used the same dataset were not included unless the corresponding contributions are significant compared to previous studies on the same data.

\item[\textbf{IC6: Timeliness --}]We observed that certain studies in smartphone sensing carried out in the first decade of the century are not necessarily relevant in today's context. Hence, to make sure that the discussion presented here are timely, we review studies from the last decade (2010 - 2020). This also ensured that the research was conducted in a time period when smartphone usage among young adults in the world was high, and that smartphone sensing related studies appeared in different sub-domains. 

\item[\textbf{IC7: Language --}]We only considered papers written in English.

\item[\textbf{IC8: Number of Participants --}]We used 10 as the minimum number of participants for a study to be considered in this review.

\end{itemize}{}

The approach we used to search and select papers for this review was the following:

\begin{itemize}[wide, labelwidth=!, labelindent=0pt]

\item[\textbf{Search:}]We used Google Scholar, PubMed, and Scopus to search for studies with different search terms related to the domain such as mobile phone sensing, smartphone sensing, smartphone sensing health, mobile sensing well-being, health young adults, etc.

\item[\textbf{Screening:}]We screened titles, abstracts, and contributions of more than 400 articles.

\item[\textbf{Eligibility:}]We used IC1-IC8 to check studies for eligibility. This process left us with 26 studies for the primary analysis. 

\end{itemize}{}

\textcolor{red}{It should be noted that in cases where certain details were not clear, we contacted the original authors to ensure that we include correct information in this review. Moreover, this study is not a systematic review, but a literature review that does not attempt to answer any specific research question \footnote{\url{https://guides.libraries.psu.edu/ld.php?content_id=36146097}}.}

\textcolor{red}{In Section~\ref{sec:organization}, we lay out the organization of this study, and in Section~\ref{study_summary}, we provide a summary regarding the selected set of papers. Further, under Data Perspective, in Section~\ref{sec:implicitsensing} and Section~\ref{sec:explicitsensing}, passive sensing and self-report modalities respectively are discussed from a human science perspective. Then, in Section~\ref{sec:system_perspective}, we examine the System Perspective. Finally, in Section~\ref{sec:discussion}, we discuss regarding insights and directions for future work.}
\section{Organization of the Review}\label{sec:organization}

\begin{figure*}[htb]
    \centering
    \includegraphics[width=\textwidth]{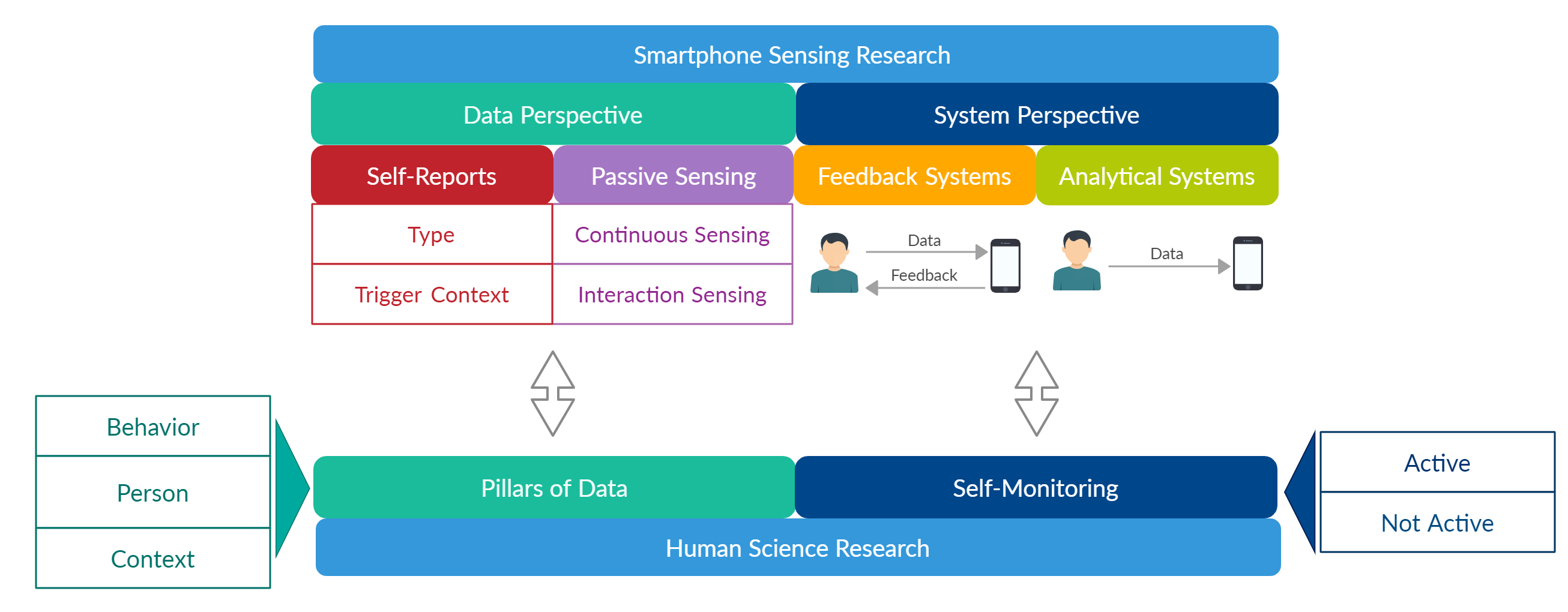}
    \caption{Taxonomy of Smartphone Sensing Studies for the Well-being of Young Adults from a Human Science Perspective}
    \label{fig:taxonomy}
\end{figure*}

\subsection{The Human Science Perspective}\label{subsubsec:behavior_modelling_theory}

\subsubsection{Pillars of Data for Smartphone Sensing.}

For purposes of organization of the review, we first borrow from Lewin’s Field Theory \cite{Lewin1936}. This classic model proposed that human behavior is affected by personal and environmental factors, broadly expressed as $B=f(P,E)$, where $B$, $P$, and $E$ stand for behavior, person, and environment, respectively. While oversimplifying the complexity and nuances of all the involved processes, this model allows for an organization of the mobile sensing work reviewed here. Descriptions of the three terms are given below.

\begin{itemize}[wide, labelwidth=!, labelindent=0pt]

\item[\textbf{Behavior:}] Changes in the life space of an individual as a result from changes in either the environment or the person. Hence, behavior represents activities and actions performed by people.

\item[\textbf{Person:}] This aspect is defined according to: (a) attributes, traits and characteristics of the individual such as beliefs, abilities, values, emotions, needs, and other person aspects; (b) static characteristics that intrinsically affect such traits like age, sex, bodily characteristics (height, weight), etc.; and (c) individuals perception of the context that they are in.

\item[\textbf{Environment:}] Conscious as well as unconscious entities in a human's environment or in other terms, the objective context at which the person perceives and acts. 

\end{itemize}{}

Drawing motivation from Lewin's field theory, we consider three ``pillars of data'' to study smartphone sensing research for the well-being of young adults, namely (1) \textbf{Behavior (B)}, (2) \textbf{Person (P)}, and (3) \textbf{Context (C)}. We mapped \textit{environment} from Lewin's theory into context data pillar in the context of smartphone sensing. We mapped {person} to P to represent the physical, mental, and social aspects of people in addition to attributes such as age, gender, and this mapping allows us to represent sensors and data sources in smartphone sensing literature as proxies to different pillars of data. Hence, we use these pillars of data in the context of smartphone sensing as a framework throughout this paper. This enables us to analyze studies in the domain using a well-grounded framework, where all the sensing data collected during the study falling into one or more of the pillars. As an example, data from the accelerometer sensor of a smartphone can be taken as a proxy regarding physical activity done by an individual, hence falling under the behavior pillar. Moreover, location coordinates fall under context pillar. If they are further processed to obtain a semantic meaning, location can be used to determine travel patterns of the users which can be categorized under behavior pillar, because changes in location could also mean a distance traveled by users, hence a behavior. Further, a self-report could reflect the emotion of a user, and it belongs to the person pillar. Hence, if a study uses features generated from accelerometer and location sensors to infer/analyze emotions of a person (target domain of the study), that reflects the use of behavior and context features to infer an attribute belonging to the person pillar. As explained above, various features generated from a single sensor might provide data regarding one or more pillars depending on the way data is processed and features are engineered. Hence, we use these three pillars of data to study the current body of research from the data perspective as explained in Section~\ref{subsubsec:data_perspective}. 

\subsubsection{Self-Monitoring as part of Mobile Sensing}

There are many well-known theories in social and cognitive sciences regarding self-monitoring and behavioral change \cite{Snyder1974,Michie2011,Compernolle2019,SelfMoni2020,Bandura1991,Boone1977}. Generally, self-monitoring is the capacity to quantify, observe, and evaluate one's behavior to facilitate behavioral change. Michie et al. \cite{Michie2011} defined self-monitoring as keeping a record of a specified behavior as a method for changing behavior. The Social Cognitive Theory (SCT) of Self-Regulation by Bandura \cite{Bandura1991} outlined that when people observe a behavior and consequences of the behavior, they learn from the events to guide their subsequent behaviors, with the potential to enable behavioral change. In addition, this body of literature presents factors that affect the quality of self-monitoring such as informativeness, regularity, proximity, and accuracy. Drawing from the essence of these ideas, we argue that mobile sensing systems that allow users to view statistics and feedback on apps, facilitate self-monitoring, thus assisting users interested in specific behavioral changes. Hence, there are two types of mobile sensing systems, one in which self-monitoring is active; and another one in which self-monitoring is not active. By processing the behavior, person, and context related data captured via smartphones, apps with active self-monitoring allow users to monitor themselves and reflect upon their behavior using feedback given in the app. Hence, while simplifying the nuances behind complex theories, we use \textit{self-monitoring} as a concept to study the current body of research from the system perspective as explained in Section~\ref{subsubsec:system_perspective}.

\subsection{The Data Perspective}\label{subsubsec:data_perspective}

The Data Perspective considers the data flow in smartphone sensing studies. Under this perspective, we primarily consider an existing classification: \textbf{Passive Sensing} (Section~\ref{sec:implicitsensing}) and \textbf{Self-Reports} (Section~\ref{sec:explicitsensing}). However, we analyze each of the components from a new perspective using pillars of data. Further, we propose a taxonomy to segregate passive sensing into: (1) Continuous Sensing, that involves embedded sensors that acquire data (e.g. accelerometer, gyroscope, location, ambient light, etc), and (2) Interaction Sensing, that captures user interactions/usage with/of the phone without utilizing any embedded sensor (e.g. app usage, phone calls, messages, typing events, etc.).

The data collected in a study could belong to either one or more pillars mentioned in Section~\ref{subsubsec:behavior_modelling_theory}. Even though continuous and interaction sensing have been named as just passive sensing in prior literature, we make this distinction to highlight some aspects regarding how smartphone sensing studies on health and well-being can benefit young adults. 

It is important to understand how data are used in smartphone sensing. In all the studies, data from different sensors are used as proxies to different phenomena  \cite{Morshed2019} related to person, context, or behavior aspects. For example, if the sensed phenomena are activity level, brightness of the surrounding, sociability, and indoor mobility, sensing modalities could be accelerometer, ambient light sensor, messaging app usage, and WiFi access point connectivity. Hence, the data perspective of a smartphone sensing study involves obtaining data from the smartphone regarding the three pillars of data in behavioral modeling. Passive sensing directly obtains data regarding these pillars from users unobtrusively, while self-report data act as proxies and ground truth for the three pillars, and are obtained explicitly from users. Hence, as a summary, smartphone sensing studies involve obtaining data from young adults belonging to three pillars using passive sensing and self-reports in order to analyze an unknown attribute belonging to one of the pillars.

\subsection{The System Perspective}\label{subsubsec:system_perspective}
Smartphone Sensing studies involve smartphone app based systems. Using self-reports and passive sensing techniques mentioned in Section~\ref{subsubsec:data_perspective}, smartphones acquire multi-dimensional data. The majority of such systems store this information for off-the-shelf analysis that is done at the end of the deployment phase. However, some systems process these data on-device or in cloud to provide feedback to users regarding their health and well-being state, hence allowing self-monitoring. If we combine (a) data acquisition by the smartphone and (b) value-added feedback given to users, together with theories regarding self-monitoring and behavior change, it is possible to divide literature as (a) Feedback Systems, where self-monitoring is active, and (b) Analytical Systems, where self-monitoring should not be active.

\textbf{Feedback systems (FSys)} do data analysis, training or inference (either in-device or on servers; done manually or in an automated manner) during the deployment phase. These systems often use results of the data analysis or inference primarily to give users in-app feedback. Some systems provide feedback regarding user behavior, mental health, and routines in order to motivate them to use the app further \cite{LiKamWa2013,Ma2012}, develop self insight \cite{Lin2012,Rodriguez2017}, provide     behavioral intervention strategies \cite{Rabbi2015}, and for clinical interventions \cite{Farhan2016}. In other terms, this type of systems that provide feedback affect normal user behavior, and can be used for interventions as well. However, it should also be noted that not all feedback systems are used for interventions because some apps do not necessarily provide any active intervention even though they summarize sensed details in the app. A feedback system with active self-monitoring should provide app users with useful feedback, and observe how people benefit from the feedback by helping them change their behavior in a positive course. Moreover, from a behavioral science perspective, systems of this sort cannot be used to evaluate a hypothesis regarding the presence or absence of a certain attribute in people because self-monitoring is active, and hence it could create biases and behavioral change in people.

\textbf{Analytical systems (ASys)} are systems that do data analysis and inference off-the-shelf after collecting the data from the study. These systems often run with no (or less) data analysis during the deployment (with the exception being activity inference from accelerometer data \cite{Wang2018,Rabbi2015,Farhan2016,Wang2014}), and even if data analysis is done, results are not directly conveyed to app users, hence avoiding any behavioral change via self-monitoring. A ASys should not have allow self-monitoring, and hence, smartphone app users should not be motivated to change their behavior. Moreover, for an ASys, it is better if app users do not necessarily know what specific hypothesis is being tested by researchers, to reduce potential biases in the mind of study participants (who need to understand the general goals of the study as part of the informed consent process). This would help to obtain better quality data which reflect the real behavior of study participants, hence leading to better studies regarding the validation of hypotheses.

\onecolumn

\begin{center}
\scriptsize

\begin{longtable}{p{2.1cm} p{0.4cm} p{1.0cm} p{0.5cm} p{0.3cm} p{3.6cm} p{0.3cm} l l l l l}

\caption{Study Details of Smartphone Sensing Research for the Well-being of Young Adults (YA)}\\

\hline

		& 
		& 
		& 
		& 
		\multicolumn{7}{c}{\textbf{Study Details}}\\
		\cmidrule(lr){5-11}
		
		\rotatebox[origin=c]{270}{Reference} &
		\rotatebox[origin=c]{270}{Year} &
		\rotatebox[origin=c]{270}{Domain} &
		\rotatebox[origin=c]{270}{Pillar} &
		\rotatebox[origin=c]{270}{Sample Size} &
		\rotatebox[origin=c]{270}{Sample Diversity } &
		\rotatebox[origin=c]{270}{Target Audience} &
		\rotatebox[origin=c]{270}{Study Length} & 
		\rotatebox[origin=c]{270}{Feedback} &
		\rotatebox[origin=c]{270}{Experts/Theory} &
		\rotatebox[origin=c]{270}{Region}
		\\ 

\endfirsthead

{\tablename\ \thetable\ \textit{Continued..}} \\ 
\\
\hline
        & 
		& 
		& 
		& 
		\multicolumn{7}{c}{\textbf{Study Details}}\\
		\cmidrule(lr){5-11}
		
		\rotatebox[origin=c]{270}{Reference} &
		\rotatebox[origin=c]{270}{Year} &
		\rotatebox[origin=c]{270}{Domain} &
		\rotatebox[origin=c]{270}{Pillar} &
		\rotatebox[origin=c]{270}{Sample Size} &
		\rotatebox[origin=c]{270}{Sample Diversity } &
		\rotatebox[origin=c]{270}{Target Audience} &
		\rotatebox[origin=c]{270}{Study Length} & 
		\rotatebox[origin=c]{270}{Feedback} &
		\rotatebox[origin=c]{270}{Experts/Theory} &
		\rotatebox[origin=c]{270}{Region}
		\\ 

\hline 
\endhead
\hline 
\multicolumn{11}{r}{\textit{Continued on next page}} \\
\endfoot
\hline
\endlastfoot

\midrule

		\thead[l]{Rachuri et al. \cite{Rachuri2010}\\EmotionSense} &
		2010 &
		Emotion&
		P &
		18&
		\thead[l]{-Members of CS Department}&
		All&
		10 days&
		&
		\checkmark&
		UK
		\\

		\midrule
		\thead[l]{Moturu et al. \cite{Moturu2011}\\Funf} &
		2011 &
		\thead[l]{Well\\being} &
		P,B &
		54&
		\thead[l]{-Graduate Students and Spouses}&
		All&
		1 month&
		&
		\checkmark&
		USA
		\\ 
		
		\midrule
		\thead[l]{Lu et al. \cite{Lu2012}\\StressSense} &
		2012 &
		Stress&
		P &
		14&
		\thead[l]{-University students\\-4 Men and 10 Women \\-13 Undergrads and 1 PhD} &
		All &
	    \thead[l]{Job\\Interviews}&
		&
		\checkmark &
		Switzerland
		\\

		\midrule
		\thead[l]{Lin et al. \cite{Lin2012}\\BeWell+} &
		2012 &
		\thead[l]{Well\\being}\footnote{Physical Activity, Sleep, Sociability}&
		B &
		27&
		\thead[l]{-Aged 21-37\\-16 Men and 11 Women\\-9\% CS Department, 34\% Doctors and \\Medic. Researchers and 57\% Grad. \\Students}&
		All&
		19 days&
		\checkmark&
		\checkmark&
		USA
		\\

		\midrule
		\thead[l]{Ma et al. \cite{Ma2012}\\MoodMiner} &
		2012 &
		Mood &
		P &
		15&
		\thead[l]{-University Students and Teachers, \\-Urban White Collar Workers}&
		All&
		30 days&
		\checkmark&
		\checkmark&
		China
		\\

		\midrule
		\thead[l]{LiKamWa et al. \cite{LiKamWa2013}\\Moodscope} &
		2013 &
		Mood&
		P &
		32&
		\thead[l]{-24 students and 8 other occupations \\-25 from China, 7 from USA}&
		All&
		2 months&
		\checkmark&
		\checkmark&
		China, USA
		\\

		\midrule
		Sano et al. \cite{Sano2013} &
		2013 &
		Stress &
		P &
		18&
		\thead[l]{-Average age 28\\-15 Men and 3 Women}&
		All&
		5 days&
		&
		\checkmark&
		USA
		\\

		\midrule
		\thead[l]{Bogomolov et al. \cite{Bogomolov2013}} &
		2013 &
		Emotion \footnote{Happiness} &
		P &
		117&
		\thead[l]{-University (Grad) Students\\-16 countries}&
		YA&
		18 months&
		&
		\checkmark&
		USA
		\\

		\midrule
		\thead[l]{Wang et al. \cite{Wang2014}\\StudentLife} &
		2014 &
		\thead[l]{Well\\Being}\footnote{Stress, Sleep, Academic Performance, Sociability}&
		P,B &
		48&
		\thead[l]{-University students\\-One CS Class\\-30 Undergrads and 18 Grad Students\\-38 Men and 10 Women\\-23 Caucasians, 23 Asians and 2 Afro-\\Americans}&
		YA&
		10 weeks&
		&
		\checkmark&
		USA
		\\

		\midrule
		\thead[l]{Bogomolov et al. \cite{Bogomolov2014}} &
		2014 &
		Stress &
		P &
		111&
		\thead[l]{-University (Grad) Students\\-16 countries}&
		YA&
		7 months&
		&
		\checkmark&
		USA
		\\

				\midrule
		\thead[l]{Rabbi et al. \cite{Rabbi2015}\\My Behavior} &
		2015 &
		\thead[l]{Well\\being}&
		B &
		16&
		\thead[l]{-7 Men and 9 Women\\-4 aged 18-29, 6 aged 30-39, \\3 aged 40-49 and 3 aged >50\\-13 experienced in Food Diary\\-11 experienced in Exercise Diary}&
		All&
		14 weeks&
		\checkmark&
		\checkmark&
		USA
		\\

        \midrule
		\thead[l]{Canzian et al. \cite{Canzian2015}\\MoodTraces} &
		2015 &
		Depression &
		P &
		28&
		\thead[l]{-University population\\-15 Men and 13 Women\\-7 students, 3 PhD students, and 7 \\researchers/lecturers}&
		All&
		10 months&
		&
		\checkmark&
		UK
		\\

		\midrule
		\thead[l]{Farhan et al. \cite{Farhan2016}\\LifeRhythm} &
		2016 &
		Depression &
		P &
		79&
		\thead[l]{-University students\\-Aged 18-25\\-26.1\% Men and 73.9\% Women \\-24.6\% Asian, 5.8\% African American, \\62.3\% Caucasian, 5.8\% multiracial}&
		YA&
		4 months&
		\checkmark \footnote{A session with a clinician when user is depressed}&
		\checkmark&
		USA
		\\

		\midrule
		\thead[l]{Abdullah et al. \cite{Abdullah2016}\\CognitiveRhythms} &
		2016 &
		Alertness &
		P &
		20&
		\thead[l]{-University Students\\-7 Men and 13 Women\\-Aged 18-29}&
		YA&
		40 days&
		&
		\checkmark&
		USA
		\\

		\midrule
		\thead[l]{Huang et al. \cite{Huang2016}} &
		2016 &
		Anxiety &
		P &
		18&
		\thead[l]{-University students\\-Undergraduates in Psychology}&
		YA&
		10 days&
		&
		\checkmark&
		USA
		\\

		\midrule
		\thead[l]{Murnane et al. \cite{Murnane2016}} &
		2016 &
		\thead[l]{Alertness \footnote{Human Body Clock}}&
		P,B &
		20&
		\thead[l]{-University students\\-7 Men and 13 Women\\-Aged 18-29}&
		YA&
		40 days&
		&
		\checkmark&
		USA
		\\

		\midrule
		\thead[l]{Seto et al. \cite{Seto2016}\\CalFit Chi and Dong} &
		2016 &
		\thead[l]{Eating\\Behavior}&
		B &
		12&
		\thead[l]{-University students\\-32.3\% Men and 66.7\% Women \\-Ages 18-31\\-2 underweight, 2 overweight and \\8 otherwise}&
		YA&
		2 weeks&
		&
		\checkmark&
		China
		\\

		\midrule
		Bae et al. \cite{Bae2017} &
		2017 &
		\thead[l]{Alcohol\\Usage} &
		B &
		38&
		\thead[l]{-21 Emergency Department (ED) \\ patients and 17 Others  \\-Aged 21-28\\-50\% Men and 50\% Women}&
		YA&
		28 days&
		&
		\checkmark&
		USA
		\\

		\midrule
		\thead[l]{Rodriguez et al. \cite{Rodriguez2017}} &
		2017 &
		Mood&
		P &
		18000&
		\thead[l]{-Full time employment and University \\students\\-Majority born between 1970 and 2000}&
		All&
		3 years&
		\checkmark&
		\checkmark&
		Worldwide
		\\

			\midrule
		\thead[l]{Wang et al. \cite{Wang2018}} &
		2018 &
		Depression &
		P &
		83&
		\thead[l]{-University students\\-Undergraduate CS students\\-40 Men and 43 Women\\-26 Asian, 5 African American, \\24 Caucasian, 1 multiracial, \\and 26 not specified}&
		YA&
		18 weeks&
		&
		\checkmark&
		USA
		\\

		\midrule
		\thead[l]{Biel et al. \cite{Biel2018}\\Bites n' Bits} &
		2018 &
		\thead[l]{Eating\\Behavior}&
		B &
		122&
		\thead[l]{-University students\\-Aged 18-26}&
		YA&
		10 days&
		&
		\checkmark&
		Switzerland
		\\

		\midrule
		\thead[l]{Santani et al. \cite{Santani2018}\\DrinkSense} &
		2018 &
		\thead[l]{Alcohol\\Usage}&
		B &
		160 &
		\thead[l]{-Aged 18-25 \\-From Two Cities \\-85 Men and 74 Women}&
		YA&
		3 months&
		&
		\checkmark&
		Switzerland
		
		\\

		\midrule
		\thead[l]{Boukhechba et al. \cite{Boukhechba2018}\\DemonicSalmon} &
		2018 &
		\thead[l]{Depression\\Anxiety}&
		P &
		72&
		\thead[l]{-University students\\-35\% Men and 37\% Women \\-Aged 18-23\\-37.5\% Asian, 4.17\% African Amer-\\ican, 41.67\% Caucasian, 4.17\% Latin, \\and 12.5\% not specified/multi racial}&
		YA&
		2 weeks&
		&
		\checkmark&
		USA
		\\

		\midrule
		Gong et al. \cite{Gong2019} &
		2019 &
		Anxiety &
		P &
		52&
		\thead[l]{-University students\\-Undergraduates in Psychology\\-32\% Men and 68\% Women}&
		YA&
		2 weeks&
		&
		\checkmark&
		USA
		\\ 
		
		\midrule
		Sagbas et al. \cite{Sagbas2020} &
		2020 &
		Stress &
		P &
		46&
		\thead[l]{-University students\\-Aged 18-39\\-35 Men and 11 Women}&
		All&
		Tasks&
		&
	    &
		Turkey
		\\

		\midrule
		Obuchi et al. \cite{Obuchi2020} &
		2020 &
		Brain Functional Connectivity &
		P &
		105 &
		\thead[l]{-University students\\-Avg. Age 18.2\\-35 Men and 11 Women\\-56\% Caucasian, 21\% Asian, 16\% \\multiracial, 3\% African, 1\% Hispanic}&
		YA&
		79 days&
		&
	    \checkmark &
		USA
		\\

\label{tab:details}		
\end{longtable}
\end{center}

\twocolumn

\section{Summary of studies}\label{study_summary}
We analyzed 26 studies, and summarized the study details under different areas as seen in Table~\ref{tab:details}. All the studies have been arranged in chronological order to understand the evolution of the body of literature. The domain column represents the target area of analysis in each of the studies. We have segregated the domain of each individual study based on the three pillar categorization under the column ``pillar''. However, in all the studies, even though the target variable belonged to one specific pillar, data belonging to other pillars have been collected using passive sensing and self-reports in order to find relationships with or to infer the target variable. As demonstrated by the evolution through the last decade, the majority of the studies are around the person pillar \cite{Rachuri2010,Lu2012,Ma2012,LiKamWa2013,Sano2013,Bogomolov2013,Bogomolov2014,Canzian2015,Farhan2016,Abdullah2016}. This could be because many of the real-world issues faced by young adults such as stress, depression, and emotional instability are in-fact socio-psychological. Before 2016, there were studies that considered both person and behavior pillars in terms of overall well-being of young adults \cite{Lin2012,Rabbi2015,Wang2014,Moturu2011}, and lately, there are studies that primarily focus on the behavior pillar \cite{Seto2016,Murnane2016,Biel2018,Bae2017,Santani2018,Abdullah2014} that are in domains such as eating behavior, alcohol consumption, and circadian behavior.

Sample size is a key design choice in all of the studies that deal with data analysis. In order to make a statistically significant conclusion, any study should have a sufficient sample size, hence why we considered studies with at least ten participants in this review. Considering that the studies discussed here have targeted young adult audiences, and the majority of them used university students as their study cohort \cite{Rachuri2010,Moturu2011,Lu2012,Ma2012,Bogomolov2013,Bogomolov2014,Wang2014,Canzian2015,Farhan2016,Abdullah2016,Huang2016,Murnane2016,Seto2016,Wang2018,Biel2018,Boukhechba2018,Gong2019}. This is explainable since most of the studies were conducted within university research labs, and university students are easily accessible in that context. 

Depending on the participant recruitment policy (e.g. convenience sampling \cite{Santani2018,Rachuri2010,Canzian2015,Huang2016,Gong2019}, study administered participant recruitment \cite{Whittaker2016,Rabbi2015,Abdullah2016,Bae2017}, in-the-wild recruitment \cite{Rodriguez2017, SeaHeroQuest2019}) employed by the researchers, studies have different sample sizes as depicted in Table~\ref{tab:details}. Although there is no concrete trend in the sample size, many recent studies emphasized the need for getting bigger populations for the studies \cite{Santani2018, Biel2018}. For example, the majority of studies considered in this review, that were published on or after 2017 have included more than 50 participants \cite{Biel2018, Santani2018, Boukhechba2018, Gong2019, Wang2018, Obuchi2020}, while the majority of studies on or before 2017 (14 out of 19) used less than 50 participants. Even though this is not an absolute figure regarding all the work done during this time, given the criteria we used to filter studies, this provides evidence of ubicomp researchers focusing on increasing sample sizes in order to make more and generalized conclusions from their studies.

Sample Diversity is a term we used to look at the constituents of the samples used for the study. The factors we considered include the type of sample (e.g. university students, researchers, patients, etc.), age group, gender, ethnic diversity, and other diversity attributes. \emph{Target Audience} is the type of end users of the system (e.g.: "All" was used when the study is for general populations, "YA" was used when the beneficiary of the study are specifically young adults). Even though very few details regarding sample diversity have been disclosed in some early studies \cite{Rachuri2010, Moturu2011}, later studies are available where a multitude of information regarding the sample diversity including demographics have been disclosed \cite{Lin2012,Wang2014,Rabbi2015,Farhan2016,Boukhechba2018,Obuchi2020}. This evolution is important in interpreting and evaluating results. For example, predicting drinking episodes of young adults has variations depending on the gender according to \cite{Santani2018}, while gender might be less crucial in identifying eating episodes \cite{Seto2016, Biel2018}. Hence, incorporating diversity data from the sample population to various analyses might help to interpret results in a more meaningful and an accurate manner.

The majority of studies done until 2013 \cite{Rachuri2010,Moturu2011,Lu2012,Lin2012,Ma2012,LiKamWa2013,Sano2013} have not particularly disclosed a specific target audience (although they used young populations in the sample in varying proportions), while a majority of studies done from 2014 onwards have been specific regarding targeting to specific audiences making it clear that their studies are valid for a specific cohort of people. When examining this in depth by analyzing sample diversity and target audiences as a whole, some observations can be made. Of all the studies which have general target audiences (e.g. "All"), some have specifically used diverse people in the sample populations \cite{Lin2012,LiKamWa2013,Rodriguez2017,Rabbi2015}; some have chosen less diverse populations \cite{Moturu2011,Sano2013}; and some have not mentioned the attributes of the sample populations \cite{Rachuri2010}. Furthermore, most of the studies were done with populations from a small set of countries namely; USA \cite{Moturu2011,Lin2012,Sano2013,Bogomolov2014,Wang2014,Bogomolov2013,Rabbi2015,Farhan2016,Obuchi2020,Obuchi2020}, Switzerland \cite{Lu2012, Biel2018, Santani2018}, China \cite{Ma2012, Seto2016}, UK \cite{Canzian2015,Rachuri2010}, and Turkey \cite{Sagbas2020}. LiKamWa et al. \cite{LiKamWa2013} carried out the study with people from both China and USA, while \cite{Rodriguez2017} is the only study in which data from several geographical regions was used. Further, the only study in which the diversity of the user location was taken into consideration in the analysis is \cite{Santani2018}, where they studied drinking behavior of young adults, observing distinct behavioral patterns in two cities. It is clear that, except for \cite{Rodriguez2017}, most of the work in this domain still lacks diversity in user populations to provide diversity-related behavioral insights in terms of geographical location. Drawing insights from these findings, a discussion regarding diversity-aware mobile computing research is presented in Section~\ref{subsec:diversity}.

On a positive note, all of the studies analyzed in this review  have relied on either the help from domain experts \cite{Rachuri2010,Lu2012,Lin2012,Wang2014,Seto2016,Rodriguez2017,Biel2018,Santani2018, Obuchi2020} or proven theories \cite{Moturu2011,LiKamWa2013,Rabbi2015,Huang2016,Sagbas2020} in passive sensing, pre and post study interviews, surveys, and focus groups. The use of behavioral science experts and theory into mobile computing research is encouraging in terms of the ability of smartphone sensing frameworks in carrying out procedures that were previously done by manual form filling, interviews, etc. 

\begin{table}[t]
\caption{Study Goals / Use Cases}
\begin{center}
\resizebox{0.5\textwidth}{!}{%
\begin{tabular}{l l l l}
	\hline
	
	Pillar &
	Domain &
	Ref. &
	Goal / Use Case
	\\
	
	\hline 
	
	\multirow{18}{*}{P}&
	\multirow{2}{*}{Emotion} &
	\cite{Rachuri2010} &
	Infer five emotions (happy, sad, neutral, etc.)
	\\
	
	&
	&
	\cite{Bogomolov2013} &
	Infer happiness 
	\\
	
	\cmidrule(lr){2-4}
	
	&
	\multirow{3}{*}{Mood} &
	\cite{Ma2012} &
	Infer three moods (displeasure, tired, tensity)
	\\
	
	&
	&
	\cite{LiKamWa2013} &
	Infer two moods (pleasure, activeness)
	\\
	
	&
	&
	\cite{Rodriguez2017} &
	Infer two moods (valence, arousal)
	\\
	
	\cmidrule(lr){2-4}
	
	&
	\multirow{4}{*}{Stress} &
	\cite{Lu2012} &
	Infer stressed or not-stressed
	\\
	
	&
	&
	\cite{Sano2013} &
	Infer high or low stress
	\\
	
	&
	&
	\cite{Bogomolov2014} &
	Infer high or low stress
	\\
	
	&
	&
	\cite{Sagbas2020} &
	Infer stressed or calm
	\\
	
	\cmidrule(lr){2-4}
	
	&
	\multirow{3}{*}{Depression} &
	\cite{Canzian2015} &
	Infer depressive state
	\\
	
	&
	&
	\cite{Farhan2016} &
	Infer clinical depression
	\\
	
	&
	&
	\cite{Wang2018} &
	Infer depressive state
	\\

	\cmidrule(lr){2-4}
	
	&
	\multirow{1}{*}{Depression \& Anxiety} &
	\cite{Boukhechba2018} &
	Analyze depressive state \& anxiety
	\\
	
	\cmidrule(lr){2-4}
	
	&
	\multirow{2}{*}{Anxiety} &
	\cite{Huang2016} &
	Infer SIAS score for anxiety
	\\
	
	&
	&
	\cite{Gong2019} &
	Analyze anxiety levels
	\\
	
	\cmidrule(lr){2-4}
	
	&
	\multirow{1}{*}{Brain Functional Connectivity} &
	\cite{Obuchi2020} &
	\thead[l]{Infer higher or lower \\vmPFC-amygdala RSFC group}
	\\
	\cmidrule(lr){2-4}
	
	&
	\multirow{1}{*}{Alertness} &
	\cite{Abdullah2014} &
	Infer alertness
	\\

	\cmidrule(lr){1-4}
	
	\multirow{3}{*}{P, B}&
	\multirow{1}{*}{Alertness} &
	\cite{Murnane2016} &
	\thead[l]{Analyze alertness and sleep \\(circadian rhythms)}
	\\
	
	\cmidrule(lr){2-4}
	
	&
	\thead[l]{Well-being \\(stress, sleep, academic work)} &
	\cite{Wang2014} &
	\thead[l]{Analyze relationship between stress, sleep,\\ and academic performance}
	\\
	
	&
	\thead[l]{Well-being \\(sleep, mood, sociability)} &
	\cite{Moturu2011} &
	\thead[l]{Analyze relationships between sleep and \\ (a) mood and (b) sociability}
	\\
	
	\cmidrule(lr){1-4}
	
	\multirow{6}{*}{B}&
	\thead[l]{Well-being \\(sleep, activity, sociability)} &
	\cite{Lin2012} &
	\thead[l]{Analyze short term behavior change \\ (sleep, activity, social interactions)}
	\\
	
	&
	\thead[l]{Well-being \\(eating, activity)} &
	\cite{Rabbi2015} &
	\thead[l]{Analyze behavior change (eating,\\ activity) using a recommendation engine}
	\\

	\cmidrule(lr){2-4}
	
	&
	\multirow{2}{*}{Eating Behavior} &
	\cite{Seto2016} &
	Analyze eating behavior
	\\
	
	&
	&
	\cite{Biel2018} &
	Infer meal or snack eating events
	\\
	
	\cmidrule(lr){2-4}
	
	&
	\multirow{2}{*}{Alcohol Usage} &
	\cite{Bae2017} &
	\thead[l]{Infer drinking episodes\\(not drinking, drinking, heavy drinking)}
	\\
	
	&
	&
	\cite{Santani2018} &
	\thead[l]{Infer drinking nights \\(drinking, not drinking)}
	\\

	\hline
\end{tabular}
}
\end{center}
\label{tab:pillars_domains}
\end{table}

Table~\ref{tab:pillars_domains} shows studies categorized based on the pillar and domain. P is the pillar that had the highest number of studies with domains such as emotion, mood, stress, depression, anxiety, brain functional connectivity and alertness. Studies included under pillar B focused on aspects such as general well-being, eating behavior, and alcohol usage. Furthermore, there are certain studies that only conducted an analysis around the focus domain and sensing data, while many studies attempted an inference task using sensing data to unobtrusively detect/characterize health and well-being related aspects. In addition, there is a wide array of well-being domains covered in this review with a more or less equal distribution of papers for each domain. To add to that, the set of studies represent well-being domains that are typically associated to young adults. Moreover, none of the studies focused on an attribute that primarily belonged to the pillar C, because in reality, C pillar does not indicate well-being attributes, but the context that affects such well-being attributes. In Section~\ref{subsec:otheryouthissues}, we discuss in more depth about the use cases discussed here and other topics related to young adults that could be explored by leveraging smartphone sensing in the future.

\section{Passive Sensing}\label{sec:implicitsensing}

In this section, we analyze the types of passive sensing in detail from the perspective of three data pillars. Table~\ref{tab:sensing} depicts which passive sensing modality is used in each study, and to which pillar of data each sensor belonged in the context of the study. In each study, each passive sensing modality can be used as a proxy to either one or more of the data pillars. As an example, in the study by Biel et al. \cite{Biel2018}, the target variable for inference was related to the eating behavior of people (inference of meal vs. snack eating events), hence it belonged to the behavior pillar. However, to infer the behavioral attribute, the study used other sensors that were captured as proxies to other data pillars (e.g.: location that belongs to context pillar, time that belongs to context pillar, sociability that belongs to person pillar, etc.). As another example, Bae et al. \cite{Bae2017} inferred the attribute alcohol consumption (behavior pillar) using accelerometer, proximity, and gyroscope sensors that were used as proxies for behavioral aspects; and location, ambient light, bluetooth, and wifi as proxies for contextual aspects. Hence, under passive sensing, we discuss the pillar of data associated for each sensor data type in the context of the study.

\begin{table*}[t]
    \centering
    \caption{Passive sensing in smartphone studies for well-being of young adults. Table depicts the pillar of data each sensor was used as a proxy for.}
    \resizebox{\textwidth}{!}{%
    \begin{tabular}{p{2cm} p{6.1cm} p{6.1cm}}
    \hline

        \textbf{Pillars of Data} &
		\textbf{Continuous Sensing} &
		\textbf{Interaction Sensing}
		\\

		\hline
		Behavior &
		
		Accelerometer: \cite{Lin2012, Ma2012, Sano2013, Wang2014, Sano2013, Rabbi2015, Farhan2016,Seto2016, Bae2017,Rodriguez2017,Wang2018,Santani2018,Boukhechba2018,Gong2019, Sagbas2020, Obuchi2020}, Proximity Sensor: \cite{Bae2017}, Location: \cite{Sano2013, Wang2014, Huang2016,Rodriguez2017,Boukhechba2018,Gong2019, Obuchi2020}, Ambient Light: \cite{Wang2014}, Audio: \cite{Rachuri2010, Lu2012, Lin2012, Wang2014, Obuchi2020}, Gyroscope: \cite{Lin2012,Bae2017, Sagbas2020}, Bluetooth: \textbf{NA}, WiFi: \cite{Rodriguez2017}, Other: \cite{Lu2012,Wang2018,Biel2018}  & 
		
		Phone Calls: \cite{Ma2012,LiKamWa2013,Sano2013,Bogomolov2013,Bogomolov2014, Bae2017,Rodriguez2017,Boukhechba2018,Gong2019}, Messages: \cite{Ma2012,LiKamWa2013,Sano2013,Bogomolov2013,Bogomolov2014, Bae2017,Rodriguez2017,Boukhechba2018,Gong2019}, Email: \cite{LiKamWa2013}, App Usage: \cite{LiKamWa2013, Murnane2016,Bae2017,Santani2018}, Browsing History \cite{LiKamWa2013}, Calendar: \textbf{NA}, Typing Events: \cite{Bae2017, Sagbas2020}, Touch Events: \cite{Sagbas2020}, Lock/Unlock or Screen On/Off Events: \cite{Sano2013, Abdullah2016,Bae2017,Wang2018,Santani2018, Obuchi2020}, Push Notifications: \textbf{NA}, Battery Events: \cite{Lin2012,Santani2018}, Other: \textbf{NA}
		\\ 
		\hline

		Person &
		
		Accelerometer: \textbf{NA}, Proximity Sensor: \textbf{NA}, Location: \textbf{NA}, Ambient Light: \textbf{NA}, Audio: \textbf{NA}, Gyroscope: \textbf{NA}, Bluetooth: \cite{Bogomolov2013}, WiFi: \textbf{NA}  & 
		
		Phone Calls: \cite{Bogomolov2013,Boukhechba2018,Gong2019}, Messages: \cite{Bogomolov2013,Boukhechba2018,Gong2019}, Email: \textbf{NA}, App Usage: \cite{LiKamWa2013, Murnane2016}, Browsing History: \textbf{NA}, Calendar: \textbf{NA}, Typing Events: \textbf{NA}, Touch Events: \textbf{NA}, Lock/Unlock or Screen On/Off Events: \textbf{NA}, Push Notifications: \textbf{NA}, Battery Events: \textbf{NA}, Other: \cite{Lu2012}
		\\
		\hline

		Context & 
		
		Accelerometer: \cite{Farhan2016, Bae2017}, Proximity Sensor: \cite{Bae2017}, Location: \cite{Ma2012, LiKamWa2013, Sano2013, Wang2014, Canzian2015, Rabbi2015, Farhan2016, Huang2016,Seto2016,Bae2017,Wang2018,Biel2018,Santani2018,Boukhechba2018,Gong2019, Obuchi2020}, Ambient Light: \cite{Bae2017, Wang2018}, Audio: \cite{Rachuri2010, Lin2012, Wang2014,Rodriguez2017,Wang2018, Obuchi2020}, Gyroscope: \cite{Bae2017}, Bluetooth: \cite{Moturu2011, Bogomolov2013, Bogomolov2014, Wang2014, Bae2017, Santani2018}, WiFi: \cite{Bae2017, Santani2018}  & 
		
		Phone Calls: \textbf{NA}, Messages: \cite{Ma2012}, Email: \textbf{NA}, App Usage: \cite{Ma2012}, Browsing History: \textbf{NA}, Calendar: \textbf{NA}, Typing Events: \cite{Bae2017}, Touch Events: \textbf{NA}, Lock/Unlock or Screen On/Off Events: \textbf{NA}, Push Notifications: \textbf{NA}, Battery Events: \cite{Bae2017}, Other: \cite{Lu2012}  

		\\
		
		\hline
\end{tabular}
}
    \label{tab:sensing}
    \vspace{-0.1 in}
\end{table*}

\subsection{Continuous Sensing}\label{subsec:continuous}
Continuous sensing does not require user interactions or system events to generate data. Typically, these sensing modalities use sensors embedded in the smartphone, and provide continuous streams of data while sampling rates and frequency of logging are determined by researchers depending on study requirements and performance considerations. 

\subsubsection{Accelerometer}\label{subsubsec:accelerometer}
Because of the habits and behaviors of young adults, they often keep smartphones with them in the pocket, bag or in hand \cite{Dey2011}. This fact is leveraged when using the accelerometer as a sensor for continuous sensing. The majority of studies discussed in this review used accelerometers to derive activity levels of users, and we observed 3 main usage types: (1) The study used raw accelerometer data to obtain various statistics (e.g. mean, median, standard deviation for sensor traces from all three axes) for different time windows \cite{Biel2018,Santani2018,Boukhechba2018}; (2) The study used external APIs such as Google Activity Recognition API \cite{GoogleActivityRecognition2019} to infer activities from accelerometer data \cite{Wang2018, Farhan2016}; and (3) The study used activity recognition algorithms developed by researchers \cite{Rabbi2015,Wang2014}. In these studies, the activity inference is done on the phone, making it suitable for real-time smartphone apps that use activity inferences to provide user feedback \cite{Rabbi2015}. 

From the perspective of data pillars, accelerometer data has often been used as a direct proxy to the behavior pillar \cite{Lin2012, Ma2012, Sano2013, Wang2014, Rabbi2015, Farhan2016, Seto2016, Bae2017, Rodriguez2017, Wang2018, Santani2018, Boukhechba2018, Gong2019}. This distinction is mainly due to the way in which accelerometer data is processed to generate features. In many studies, accelerometer data (using it as a proxy regarding the behavior pillar) has been used together with other data to infer person attributes such as stress and emotions. In some studies, they are used together with other sensors to infer other behavioral patterns such as eating or drinking behavior. 

\subsubsection{Proximity}\label{subsubsec:proximity}
This sensor measures whether anything is close to the screen of the phone, while making sure that the screen is turned off, if necessary to save battery, and to avoid accidental touches to the screen. The proximity sensor has not been used often in the literature for determining activity levels of users or in that matter, for any other use-case. However, Bae et al. \cite{Bae2017} used it to identify whether the phone is in the pocket or not, hence, obtaining data with regard to the context and the behavior. They used proximity data together with other information such as app usage (behavior) and battery events (context) to generate features related to the overall device activity of the individual (behavior). There is ample opportunity to explore how this sensor can be used in creative ways to use it as a proxy to behavior (e.g.: calling using earphones or calling from the phone, how close the phone is to the face of the user might have details regarding the eye conditions of users etc.) or person pillars.

\subsubsection{Location}\label{subsubsec:location}
The location is the most common sensor found in literature with regard to continuous sensing. Even though in some cases, the location has been used as a proxy to sociability (person) \cite{Rachuri2010}, it has been primarily used as a proxy for context or behavior pillars of data as depicted in Table~\ref{tab:sensing}. One of the most common uses of location is to derive location-based behavioral patterns of users. Techniques such as Latent Dirichlet Allocation (LDA) and Hierarchical Dirichlet Process (HDP) have been commonly used for such cases \cite{Phan2019}. Further, there are a number of features that can be derived from location such as the distance traveled \cite{Bae2017, Canzian2015}, time at home and the university \cite{Bae2017}, points of interest \cite{Bae2017, Huang2016}, entropy \cite{Bae2017}, radius of gyration \cite{Bae2017, Canzian2015}, number of different places visited \cite{Canzian2015}, etc. which makes this sensing modality versatile in terms of the features that can be generated. There are three main techniques to determine the location in smartphones: (1) GPS Sensor \cite{Biel2018, Wang2014, Canzian2015} - This sensor has been used in many studies as the primary source of location sensing. Even though the accuracy of determining the location is high, it is not power-efficient; (2) Phone Signals \cite{Rodriguez2017,Rachuri2010, Farhan2016} - Phone signal information also carries location information. Even though this technique is far less battery consuming, the accuracy of determining the location is low; and (3) WiFi Access Points \cite{Bae2017, Santani2018, Rodriguez2017} - It is possible to determine the location (moderately higher accuracy compared to phone signals) using WiFi access points. However, it is required that the WiFi connection of the phone is kept turned on. This kind of location sensing is crucial for indoor positioning and localization too.

According to the Table~\ref{tab:sensing}, location has not been directly used as a proxy to person attributes in smartphone sensing literature, even though it has been used (as a data source belonging to {context} pillar of data) together with other data types. It is known that location can affect the mood and stress levels of people \cite{Sandstrom2017}. Considering semantic labels regarding the location of users (e.g. in a restaurant, night club, home, school, etc) and associating those semantic labels to stress or emotional variables of young adults would help to use location directly as a proxy to person data. This is another avenue in which future research can be conducted (e.g. inferring whether a person would consume alcohol in the evening, considering stress levels and other factors related to locations he has been to during the day). 

\subsubsection{Ambient Light}\label{subsubsec:ambientlight}
This sensor is used in the smartphone for the purpose of adjusting screen brightness depending on the surrounding lighting condition. Hence, it allows measuring the lighting condition the user is in. Only four out of 26 studies used this sensor in their analyses. In \cite{Ma2012, Bae2017}, it was used to measure well-lit environments or the position of the phone (e.g. in the bag, pocket, on the table, etc.), hence providing data belonging to the context pillar, while in \cite{Wang2018, Lin2012}, it was used in determining sleeping patterns, providing data corresponding to the behavior pillar. While there are numerous studies \cite{Kamarulzaman2011, Smolders2014} that discuss the effect of ambient light conditions in stress, fatigue, and mental well-being, this connection has not been extensively studied in smartphone sensing literature on the well-being of young adults. It is a worthwhile problem to look into because many young adult students tend to be awake at night for longer time periods, and this might have a considerable effect on their mental well-being. 

\subsubsection{Audio}\label{subsubsec:audio}
Audio can be passively recorded given the consent of the users. This sensor is used to extract features that can be used for multiple purposes: human voice \cite{Lu2012, Rachuri2010}, background noise and the sound of the environment \cite{Rodriguez2017}, and conversations \cite{Wang2014}. Hence, studies used audio sensor (mic) to capture for varying purposes including inferring mental well-being, stress, and sociability (all of which belonging to the person pillar). In most of these studies, what audio directly measures is the {contextual} data regarding the users (e.g. background noise, social context), and they use these contextual information to infer the target variable belonging to the person pillar. Rachuri et al. \cite{Rachuri2010} used passively sensed audio as a proxy regarding the context and behavior, and used the data to infer emotions from speech, while in \cite{Rodriguez2017}, they used background noise as a proxy regarding the environmental conditions (context). Further, \emph{StressSense} \cite{Lu2012} only used audio from the microphone to infer stress, and specifically, they considered how people speak to measure stress levels of university students who are facing job interviews. \cite{Rodriguez2017, Lin2012} further discussed the use of audio for determining sleeping behavior. In \emph{StudentLife} \cite{Wang2014}, two classifiers were used for human voice and conversations in a single pipeline, with the goal of determining how long a conversation lasts. They used this as a measure of sociability and context of students. In addition, they use audio data in determining sleep patterns too. An important factor when using audio is the privacy of users. On a positive note, some studies \cite{Rodriguez2017, Rachuri2010, Wang2014} have made sure they followed certain policies such as: (1) never record audios, (2) never send audio to the cloud, and (3) never analyze speech content to make sure that the privacy of users is preserved. For people to trust audio-based smartphone apps, this kind of privacy-preserving techniques are extremely important to be incorporated. This is clearly a current problem with popular devices and systems such as Alexa and Google Home that have been shown to involve people listening to individual audio snippets without user knowledge \cite{Sarkar2019, Day2019}.

\subsubsection{Gyroscope}\label{subsubsec:gyro}
The gyroscope is used to measure the orientation of the smartphone, and is used often in mobile games, augmented reality and virtual reality related applications as part of the trio of sensors (accelerometer, gyroscope, and magnetometer) that make up the inertial measurement unit (IMU unit) \cite{Ahmad2013}. It is another sensor that has been rarely used in smartphone sensing research in general, although some activity recognition algorithms use gyroscope data in addition to accelerometer data \cite{GoogleActivityRecognition2019}. When it comes to research targeting the well-being of young adults, we came across only one study which used the capabilities of a gyroscope. Bae et al. \cite{Bae2017} used the rotation of the phone derived from the gyroscope in order to study the drinking behavior of young adults. They concluded that movement and location-based features (e.g.: accelerometer mean magnitude, maximum magnitude of rotation, radius of gyration) can contribute to determining drinking episodes. There is plenty of opportunity to test the validity of using the gyroscope sensor data as a proxy to person and context pillars. For example, can smartphone sensing studies determine phone usage behavior of people (e.g.: whether they are using the phone while keeping it on table, on the hand, or while on the bed where gyroscope readings would be significantly different from each other) and leverage these features to infer person and behavior attributes (e.g. are people using smartphone more while on bed more stressed/healthy compared people who do so while being on a table, etc.)?

\subsubsection{WiFi and Bluetooth}\label{subsubsec:wifi}
WiFi is the primary way through which most smartphone users access the internet. Even though many smartphone studies reviewed here used WiFi connections for transmitting data to servers \cite{Moturu2011, Rabbi2015}, only a few studies \cite{Ma2012, Bae2017, Santani2018, Wang2018, Rodriguez2017} used WiFi related data for a purpose other than data transmission. WiFi has been used as a mode of location sensing in all of these studies, while \cite{Ma2012, Bae2017, Santani2018, Rodriguez2017} used WiFi in a hybrid location-sensing strategy accompanied by GPS and cellphone signals as explained in Section~\ref{subsubsec:location}. \cite{Wang2014} used WiFi signals to get fine-grained indoor positions using the indoor localization system at Dartmouth College.

The bluetooth technology has been used in smartphone sensing research on young adults, primarily to understand the context of users. Several studies \cite{Do2011v2, Rachuri2010,Moturu2011,Bogomolov2014,Bogomolov2013} used bluetooth logs to infer physical proximity to others, and to determine co-location (context). Other works such as \cite{Santani2018} used features derived from bluetooth scans such as number of records, number of unique bluetooth IDs in range, and the percentage of empty bluetooth scans as general proxies to social context. Moreover, \cite{Bae2017} used bluetooth network usage and \cite{Wang2014} used bluetooth connections to transmit data from wearable bands used in their study. In this case, bluetooth is only used for connectivity rather than as a behavioral cue by itself.

If we consider continuous sensing techniques as a whole, it is visible that they have been often used to infer person attributes by taking them as proxies for context or behavior pillar data. Avenues for future research in continuous sensing are discussed in Section~\ref{sec:discussion}. 

\subsection{Interaction Sensing}\label{subsec:interactive}
Interaction sensing does not require a physical sensor to be present, even though it captures data regarding users' interaction with a smartphone. These modalities are based on the software used in the system and events triggered in the smartphone. The informativeness of the sensing modality depends on the way people use the phone. Interaction sensing modalities generate data based on events that are triggered by the smartphone user. When an event occurs, the data can be logged to data repositories. e.g.: phone calls, messages, app usage, browsing history, calendar, typing events, touch events, lock/unlock events. These data have been often used as direct proxies to either \textit{behavior} or \textit{person} attributes.

\subsubsection{Phone Calls and Messages}\label{subsubsec:phonecalls}
Several studies have put an effort to leverage call events and message events as virtual sensing modalities for person attributes of young adults. Most of the research that use these two interaction sensing techniques \cite{LiKamWa2013,Sano2013,Bogomolov2013,Bogomolov2014,Rodriguez2017} are related to the mental well-being, while one study dealt with inferring drinking episodes (behavior) \cite{Bae2017}. Some general features derived from these two types of modalities include the number of phone calls, number of messages, duration of phone calls, and length of message \cite{LiKamWa2013,Sano2013,Bogomolov2013,Bogomolov2014,Rodriguez2017}, while some studies have delved in depth by analyzing types of features derived from these modalities.

\cite{Sano2013,Bogomolov2014,Bogomolov2013,Bae2017} used incoming and outgoing calls as different features in their studies, while Sano et al. \cite{Sano2013} found that fewer SMSs sent out correlated with higher stress levels in young adults. Bae et al. \cite{Bae2017} used features such as speed of typing, number of emojis in messages, types of emojis, number of unique conversations and they found positive relationships between features such as number of deletions, number of key-press insertions, average time between key-presses on one hand, and drinking episodes on the other hand. In a study about the anxiety of college students, Gong et al. \cite{Gong2019} used accelerometer and location data in and around phone calls or messages, and were able to infer anxiety episodes using these features. Moreover, Murnane et al. \cite{Murnane2016} used phone calls and SMSs as app usage events instead of isolated call or message events and measured phone app usage time and SMS app usage time instead of the usual event log details. Since most smartphones include phone and SMS features in the form of apps, this strategy could be said to be beneficial, while the concrete added value of these measurements over usual measures such as the number of calls and messages needs further research. There is also the question of whether messages or voice calls are useful features in the current day and age, especially among young adults who frequently use apps such as Whatsapp, Skype, Viber and Facetime for communication purposes. Hence, this needs further research.  

\subsubsection{App Usage and Browsing History}\label{subsubsec:appusage}
Even though young adults tend to use smartphone apps regularly for different usage scenarios, and a body of research regarding smartphone app usage behaviors in other research domains exist \cite{Mathur2017,Bohmer2011,Do2011,Do2010}, this aspect has not been used extensively in smartphone sensing related research targeting young adults with few exceptions. We found only 3 studies \cite{LiKamWa2013,Murnane2016,Santani2018} that directly leverage app usage behavior for their analyses by generating behavioral features from app usage.
 
There are different types of apps used by people for different purposes, and usage patterns of these apps can portray a picture regarding the daily routines (behavior) as per Murnane et al. \cite{Murnane2016}. In their work, they emphasize the importance of app usage features related to apps running in the foreground instead of the total app running times (including when running in the background), as the former type of features are more informative. Further, they categorize smartphone apps into entertainment, games, communication, browsing, etc. to gain a fine-grained understanding regarding app usage behavior and circadian rhythms of people. Santani et al. \cite{Santani2018} and LiKamWa et al. \cite{LiKamWa2013} used similar approaches for their studies in the domains of alcohol usage (behavior) inference and mood (person) inference. In both studies, they consider the most used apps and categorize them into app groups using different criteria. Further, they consider app launch events and app usage duration in their analyses. Moreover, \cite{Santani2018} reports that they were able to discriminate drinking episodes with an accuracy of 61\% using app usage alone. The most frequently used app in their study was Whatsapp (for messaging and communication), while \cite{Murnane2016} found out that communication apps are frequently used between 9 a.m. - 9 p.m., whereas social media apps are the top used category before sleeping and after waking up (after 9 p.m. and before 9 a.m.). As depicted in Table~\ref{tab:sensing}, app usage has not been used in the context of contextual pillar that much. Some interesting avenues to explore are: (a) Driven by the fact that majority of young adults use mobile fitness applications \cite{Comstock2018}, explore whether smartphone fitness app usage does have correlations with the real physical activities or fitness, hence making it possible to use app behavior as a proxy for behavior pillar; and (b) whether the apps that are used at a particular moment are correlated with the context/location of a user. e.g.: whether people use the same kind of apps while they are traveling, eating, in class, in a bar, etc. High correlations might enable us to directly use highly used app categories as a proxy to the context of users.

Browsing History is another modality that can be derived from smartphones. Even though internet browsers have been used as an application type in certain studies \cite{Santani2018, Murnane2016}, they have not particularly analyzed the browsing history-related features in connection to the problem domain in which they have worked on. On the other hand, LiKamWa et al. \cite{LiKamWa2013} used the unique website domains visited as a feature (behavior) in their model to infer the mood of individuals (person). As depicted in Table~\ref{tab:sensing}, there are plenty of opportunities to use internet browsing as proxies for all three pillars. An interesting area to look forward would be to use a categorization of websites (similar to app categorizations) and to understand person attributes. E.g.: number of visits to websites related to education, knowledge discovery such as StackOverflow, Google Scholar might have correlations with the intellectual capabilities of individuals and mental well-being. 

\subsubsection{Phone Usage Events}\label{subsubsec:phoneusage}
This relates to sensing modalities and event triggers related to phone usage such as typing, touching, screen on and off, locking and unlocking the phone, battery events. Sano et al. \cite{Sano2013} demonstrated in their study that screen on/off times (behavior) have correlations with stress levels (person) of young adults. They derived a set of features, out of which standard deviation of the percentage of the screen on events between 6 pm - 9 pm demonstrated correlations with stress, and also worked with 75\% accuracy in predicting stress events. Further, Abdulla et al. \cite{Abdullah2016} used the screen on/off related features (behavior) to determine the overall alertness of university students (behavior/person). They also used touch events (behavior) to determine alertness more as a mechanism of self-reporting (discussed in Section~\ref{sec:explicitsensing}) in accordance with a standard testing mechanism known as Psychomotor Vigilance Task (PVT). 

Bae et al. \cite{Bae2017}, who analyzed message typing behavior, also paid attention to other typing related dynamics. They showed that there is a correlation between features related to general and keypress typing (behavior) with drinking episodes (behavior). They further utilized screen on/off times to determine phone usage behavior and showed correlations between those metrics and drinking episodes. Moreover, they concluded that screen unlocks per minute is likely to be lower when a user is drinking in comparison to non-drinking episodes. They used battery charging event-related metrics in their analysis too. In another study on alcohol drinking episode prediction, Santani et al. \cite{Santani2018} used screen on/off related features to determine phone usage. They concluded that the percentage of screen on time (behavior) is negatively correlated to alcohol consumption (behavior). Further, they were able to predict drinking episodes with 61.1\% using only phone usage behavior. Wang et al. \cite{Wang2018} used phone usage in classrooms (behavior) as a proxy for lack of concentration (person) in university students. The study was built on previous literature regarding smartphone usage and concentration \cite{Park2012,Elhai2017}, while they specifically measured smartphone usage events in classrooms, dorms, and throughout the day as a whole using lock and unlock events of the phone.

As a whole, in terms of generating features belonging to pillars of data, interaction sensing modalities have been extensively used to generate features as proxies to person and behavior attributes. As some social science research shows \cite{Lepp2013}, there are connections between phone usage, physical activity, and sedentary behavior in young adults. Further research in the smartphone sensing domain is required to validate these claims, and to directly generate features that can be used as proxies for attributes belonging to the behavior pillar.

\section{Self-Reports}\label{sec:explicitsensing}

\begin{table*}[t]
    \centering
    \caption{Self-Report Data in Smartphone Sensing Studies for Well-being of Young Adults that Belong to Different Pillars of Data.}
    \resizebox{\textwidth}{!}{%
    \begin{tabular}{p{2cm} p{3.9cm} p{3.9cm} p{3.9cm}}
    \hline

        \textbf{Pillars of Data} &
		\textbf{Type} &
		\textbf{Trigger Context} &
		\textbf{Pre and Post Deployment} 
		\\

		\hline
		
		Behavior & 
		
		Structured QnA: \cite{Rachuri2010, Moturu2011, Lin2012, Sano2013, Wang2014, Rabbi2015, Abdullah2016,Bae2017,Santani2018, Biel2018}, Pictures from Camera: \cite{Rabbi2015,Biel2018}, Videos from Camera: \textbf{NA}, Audio: \textbf{NA}, Diary: \cite{Murnane2016, Lin2012}, Other: \cite{Sagbas2020} & 
		
		In-situ: \cite{Lin2012, Abdullah2016,Biel2018,Santani2018,Rabbi2015, Sagbas2020}, Retrospective: \cite{Rachuri2010, Moturu2011, Murnane2016,Bae2017,Biel2018,Santani2018,Wang2014, Rabbi2015}, Self-Initiated: \cite{Rachuri2010, Moturu2011, Lin2012}, Reminders: \cite{Murnane2016,Bae2017,Biel2018,Santani2018, Wang2014} & 
		
		Interviews/Focus Groups: \cite{LiKamWa2013, Murnane2016,Bae2017,Boukhechba2018}, Filling Survey: \cite{Sano2013, Rodriguez2017}, Other: \textbf{NA}
		\\
		\hline

		Person & 
		
		Structured QnA: \cite{Rachuri2010, Moturu2011, Sano2013, Bogomolov2013, Ma2012, Bogomolov2014, Wang2014, Canzian2015, Rabbi2015, Farhan2016, Abdullah2016, Seto2016, Wang2018, Boukhechba2018}, Pictures from Camera: \textbf{NA}, Videos from Camera: \textbf{NA}, Audio: \textbf{NA}, Diary: \textbf{NA}, Other: \cite{LiKamWa2013, Wang2014, Rabbi2015, Abdullah2016, Murnane2016, Rodriguez2017} & 
		
		In-situ: \cite{Ma2012, LiKamWa2013, Wang2014, Abdullah2016, Murnane2016, Seto2016, Rodriguez2017, Wang2018, Boukhechba2018}, Retrospective: \cite{Rachuri2010, Moturu2011, Bogomolov2013, Bogomolov2014, Wang2014, Canzian2015, Seto2016, Boukhechba2018}, Self-Initiated: \cite{Rachuri2010, Moturu2011, Ma2012, Bogomolov2013, Bogomolov2014, Wang2014, Murnane2016, Rodriguez2017}, Reminders: \cite{LiKamWa2013, Canzian2015, Farhan2016, Seto2016, Boukhechba2018} & 
		
		Interviews/Focus Groups \cite{LiKamWa2013, Farhan2016, Murnane2016}, Filling Survey \cite{Sano2013, Bogomolov2014, Bogomolov2013, Wang2014, Huang2016, Rodriguez2017, Wang2018, Gong2019}, Other: \cite{Obuchi2020}
		\\
		\hline

		Context & 
		
		Structured QnA: \cite{Rachuri2010, Sano2013, Wang2014, Seto2016, Bae2017, Biel2018, Santani2018}, Pictures from Camera: \cite{Rabbi2015, Biel2018}, Videos from Camera: \cite{Seto2016}, Audio: \cite{Seto2016}, Diary: \cite{Lin2012}, Other: \textbf{NA} & 
		
		In-situ: \cite{Lin2012, Seto2016, Biel2018, Santani2018}, Retrospective: \cite{Rachuri2010, Seto2016, Bae2017, Biel2018, Santani2018}, Self-Initiated: \cite{Rachuri2010, Lin2012, Biel2018}, Reminders: \cite{Bae2017, Santani2018} & 
		
		Interviews/Focus Groups: \textbf{NA}, Filling Survey: \cite{Rodriguez2017}, Other: \textbf{NA}
		
		\\

		\hline
\end{tabular}}
    \label{tab:reports}
    \vspace{-0.1 in}
\end{table*}

In this section, we discuss regarding data captured via self-reports used in smartphone sensing studies for the well-being of young adults as given in Table~\ref{tab:reports}. While passive sensing data are used as proxies for attributes of different pillars, self-report data are often used as ground truth events \cite{Pejovic2016v2}, and they too can belong to either one or more pillars. There are some studies in which ground truth data are obtained manually, before or after the deployment phase (pre/post study), while many studies use the ubiquity of smartphones in obtaining real-time self reports in the form of Ecological Momentary Assessments (EMA) and Questionnaires \cite{Rodriguez2017, Santani2018, Biel2018}. In Table~\ref{tab:reports}, we have divided self-reports into two sections: (1) \emph{Type} - What is the type of method used to capture self-report data from the user; and (2) \emph{Trigger Context} - In what kind of a situation/context the self-reporting is captured, and whether users are reminded regarding self reports. Further, we consider whether filled surveys, interviews or focus groups are involved in studies, to collect additional data in any of the three study phases. We analyze all those aspects in terms of the pillars of data.

\subsection{Types of Self-Reports}\label{subsec:typeofselfreports}
\subsubsection{Questionnaires, EMA and Diaries}\label{subsubsec:questions_ema} Structured questionnaires \cite{Moturu2011,Ma2012,LiKamWa2013,Sano2013,Rabbi2015,Canzian2015} and Ecological Momentary Assessments \cite{Shiffman2008} (EMA) \cite{Wang2014,Abdullah2016,Seto2016} are the most common ways of collecting self reports from users. With these questionnaires, users were given multiple choice questions, and yes or no questions depending on the type of ground truth which needs to be obtained. Further, most studies used standard questionnaires or techniques borrowed from social psychology and health related literature in order to make sure that the questions asked in the self reporting phase are of a solid foundation \cite{Moturu2011,Ma2012,Sano2013,Wang2014,Canzian2015,Farhan2016,Huang2016,Murnane2016}. Some studies have even used the expertise of personals in respective health, well-being domain in order to create standardized questionnaires \cite{Rachuri2010,Lu2012,Lin2012,Wang2014,Seto2016,Rodriguez2017,Biel2018,Santani2018}. Structured QnA, EMA were commonly used to obtain ground truth related to person pillar, using both in-situ and retrospective approaches. As given in Table~\ref{tab:reports}, questionnaires and EMA are the primary types of self-reports used in the body of literature, and these self-report techniques have been used to capture features regarding all three pillars of data. Rachuri et al. \cite{Rachuri2010} used a diary to explicitly report emotions (person) in their study, and it was mentioned that a group of psychologists and domain experts were involved in the study design, while \cite{Murnane2016} used a sleep diary for sleep related (behavior) data collection in their study. Moreover, in most studies, self-reports were used to obtain data regarding contextual aspects, as many nitty gritty details regarding the context can not be captured via techniques available in smartphones currently. For example, Santani et al. \cite{Santani2018} and Biel et al. \cite{Biel2018} in their behavioral studies regarding drinking and eating respectively, collected data regarding social context (context) and concurrent activities (behavior) that are done while drinking/eating. 

\subsubsection{Photo, Audio and Video}\label{subsubsec:photos_audio_video} 
These modalities are traditionally discussed in multimedia research, and is mainly focused on the content of the multimedia. With the emergence of ubiquitous smartphones that have become primary cameras for many individuals world-wide \cite{Cakebread2017}, it has become an important area of discussion in ubiquitous computing as well. While multimedia and computer vision researchers are focused on generating different techniques to obtain features from photos and audio, ubiquitous computing domain has focused on using these contextual and behavioral features for pervasive applications to sense humans. Further, it should be noted that audio has been used as a passive sensing modality in many studies as described in Section~\ref{sec:implicitsensing}. In this case, we consider instances where audio data has been reported by users. 

In a study regarding the eating behavior of university students, Biel et al. \cite{Biel2018} used photographs taken from the smartphone as a way of self reporting. Users were asked to take a picture of the meal/snack they are taking and upload the picture together with the structured questionnaire that they had to fill in. Further, in another eating behavior related study, Seto et al. \cite{Seto2016} used videos with voice annotations in order to determine portion sizes and food groups of meals. Some of the considerations \cite{Biel2018, Seto2016} mentioned about using photos and videos are the high data usage and the possibility of using crowd-sourcing to annotate videos/photos. Interestingly, even-though audio has been used as a passive sensing modality (see Section~\ref{subsubsec:audio}), we did not find studies where audio has been used for self reporting except for audio annotations of videos in \cite{Seto2016}.

Another important aspect with regard to photo and video based self-reports is that it is highly likely to be free of human bias compared to answering questions or using EMAs. Moreover, from the perspective of three pillars, photos, videos, and audio were not used extensively as proxies to {person} attributes. While it is obvious given that the sensing technique uses the smartphone camera, innovative mechanisms might be able to tackle such cases. For example, young adults are obsessed with taking selfies, and leveraging this by asking young adults to upload selfies with their reports might possibly help determine the contextual aspects (e.g. light levels of the room, what is present in the background, etc.) and person aspects (e.g. facial expressions of the users). A more adventurous envisaged application might use facial features such as beard, hair, etc. to generate interesting novel features (e.g. growing beard, cutting hair, emotions, etc.) that could be useful in generating fine-grained behavioral or socio-psychological (person) patterns over time. 

\subsection{Trigger Context of Self-Reports}\label{subsec:triggercontext}
Trigger context is the context at which the smartphone users initiate the self reporting procedure, and the context of the initiation it self. We have segregated it into 4 main sections as In-situ, Retrospective, Self-Initiated, and Reminders. It should be noted that some studies have not disclosed specific details regarding certain trigger contexts. Hence, we gave our best effort to categorize the studies from the details they have provided. 

\subsubsection{In-situ}\label{subsubsec:insitu} In-situ experience sampling is where the smartphone users report details regarding the current situation they are in \cite{LiKamWa2013,Wang2014,Abdullah2016}. This is beneficial for understanding the user behavior, and to obtain accurate ground truth data. Some example questions that can be asked are "How do you feel now?" and "What are you eating now?". LiKamWa et al. \cite{LiKamWa2013} used in-situ self-reports to obtain the mood of young adults (person) with an interface they created based on Circumplex Mood Model and Positive and Negative Affect Scale (PANAS). They designed the study such that users reported their current mood four times a day, each report three hours apart. Wang et al. \cite{Wang2014} used EMAs several times per day to obtain details regarding depression and stress (both under person pillar) of university students. In another study regarding alertness of people (behavior/person), Abdulla et al. \cite{Abdullah2016} used Psychomotor Vigilant Task (PVT) several times per day to obtain the in-situ alertness of smartphone users. In another study regarding the eating behavior of university students (behavior), Biel et al. \cite{Biel2018} used in-situ reporting of meals and snacks with images taken from the phone which they used to obtain portion sizes, nutritional values and also to enforce compliance. Wang et al.\cite{Wang2014} used in-situ self reporting in an interesting manner where they changed the frequency of self reports gathered from users depending on the time period. For example, they mentioned that they used more frequent self reports to obtain more stress related details during time periods closer to assignment deadlines. Hence, in-situ self reporting has been used reasonably well to capture data regarding the person pillar (e.g.: mood, emotion, feelings, and alertness) and behavior pillar (e.g.: eating, drinking, and sleeping habits), hence avoiding the recall bias, because data are captured then and there. Moreover, there is a considerable increase in the use of in-situ experience sampling techniques over time in the set of studies within our scope.

\subsubsection{Retrospective}\label{subsubsec:retrospective} The retrospective trigger context considers situations where the self-reporting is done retrospectively, and not regarding the current situation. Some examples for questions of this nature are "How was your mood in the morning today?", "What did you eat for lunch", and "How was your sleep last night?". This type of self-reports are common in most studies according to our findings. Bae et al. \cite{Bae2017} used retrospective self-reporting with high completion rates in their study on alcohol usage patterns of young adults. They also reported from one of their previous studies that using hourly EMAs to obtain alcohol consumption had low reporting rates. Some studies also used end of the day retrospective surveys to obtain details from users regarding certain details about the whole day \cite{Moturu2011,Sano2013,Boukhechba2018}. Further, \cite{Santani2018,Biel2018} used forgotten drinks surveys and forgotten meal/snack surveys in order to obtain retrospective details from university students in cases where they have forgotten to complete the in-situ reporting. While retrospective techniques might be convenient for smartphone users because they are not disturbed throughout the day, it does not allow researchers to collect real-time data. Another drawback of this technique is that it might lead to recall bias. 

\subsubsection{Self-Initiated}\label{subsubsec:selfinitiated} The self-initiated trigger context refers to instances where users are supposed to initiate the self report them selves without any reminders. Self-Initiated surveys can be both \emph{in-situ} or \emph{retrospective}. In Table~\ref{tab:reports}, there are few main types of studies which were categorized under \emph{Self-Initiated}. Namely, studies that use self reporting mechanisms and have not mentioned details regarding user reminders/notifications \cite{Rachuri2010}; studies that used both reminders and self-initiated reporting options \cite{Santani2018, Bogomolov2014, LiKamWa2013, Rodriguez2017}; studies that used notification mechanisms when self-initiated reporting is not done \cite{LiKamWa2013, Rodriguez2017}; and studies that require self-reports only when an event occurs (e.g.: eating, drinking, etc.) \cite{Biel2018, Santani2018}. 

\subsubsection{Reminders}\label{subsubsec:reminders} Reminders, or in the context of smartphones, \emph{Push Notifications} are used in studies to remind young adults regarding self-reports. Reminders can be useful in increasing the self-reporting rates and compliance \cite{LiKamWa2013, Santani2018}. There are studies in which notifications have been sent several times per day in order to collect in-situ EMAs \cite{LiKamWa2013} while in some studies, phone notifications have been used to remind users regarding retrospective surveys. Phone notifications might help users in some cases when it is difficult for them to keep track of the self-reports, and keep completing them. On the other hand, too many notifications might be disturbing for users and this aspect should be considered when choosing notification strategies for real-world deployments \cite{mehrotra2017}. However, studies regarding young adults \cite{Warc2015} have shown that young adults respond to location based push notifications in a positive manner. Leveraging this behavior of young adults could benefit smartphone sensing studies, specially since push notifications have not been discussed thoroughly as a proxy to any of the three pillars as given in Table~\ref{tab:sensing}. With preliminary findings suggesting the effect of push notifications in the productivity and stress (person) of people \cite{Pielot2017}, \cite{PushNotifications2018}, \cite{PushNotifications2019}, \cite{Morrison2017}, it is worth being explored more on how people respond to push notifications, and also how push notifications affect the behavioral and socio-psychological (person) state of individuals.

\section{System Perspective}\label{sec:system_perspective}

In social science and clinical experiments where researchers try to reduce bias imposed on study participants, they use various techniques to make sure that they do not inflict behavioral change (e.g. in randomized controlled trials, a group does not know exactly what they are tested for). This ensures that researchers are able collect unbiased and accurate data from human subjects regarding their normal behavior. When considering smartphone sensing studies by the essence of self-monitoring, we are able to segregate all the studies into two types: (1) Analytical Systems and (2) Feedback Systems.

\begin{table}[t]
    \centering
    \caption{Studies from the System Perspective}
    \resizebox{0.48\textwidth}{!}{%
    \begin{tabular}{p{6.7cm}}
    \hline

        \\ 
        Analytical Systems \cite{Rachuri2010}, \cite{Moturu2011}, \cite{Lu2012}, \cite{Sano2013}, \cite{Bogomolov2013}, \cite{Wang2014}, \cite{Bogomolov2014}, \cite{Canzian2015}, \cite{Farhan2016}, \cite{Abdullah2016}, \cite{Huang2016}, \cite{Murnane2016}, \cite{Seto2016}, \cite{Bae2017}, \cite{Wang2018}, \cite{Biel2018}, \cite{Santani2018}, \cite{Boukhechba2018}, \cite{Gong2019}, \cite{Obuchi2020}, \cite{Sagbas2020}
		\\
		\\ 
		Feedback Systems \cite{Lin2012}, \cite{Ma2012}, \cite{LiKamWa2013}, \cite{Rabbi2015}, \cite{Rodriguez2017}
		\\ 
		\\ 
		\hline 
    \end{tabular}}
    \label{tab:feedback}
    \vspace{-0.1 in}
\end{table}

\subsection{Analytical Systems}\label{subsubsec:analyticalsystems}
Analytical Systems run primarily for the purpose of data collection while no (or less) analysis, training or inference is done during the deployment phase. Another common feature of these systems is that they do not provide any feedback to the user regarding health, well-being, or behavioral routines, hence not assisting with self-monitoring. All the data collected during the study, are then used for offline analysis when the study is finished. Majority of studies (except for the once mentioned in Section~\ref{subsubsec:feedbacksystems}) which target the well-being of young adults, fall under this system perspective. Some examples for this type of systems are studies by Santani et al. \cite{Santani2018}, Biel et al. \cite{Biel2018}, Wang et al. \cite{Wang2018} etc. Further, there is a visible trend in studies in-terms of decreasing number of FSys and increasing number of ASys. 

The goal of analytical systems is to test for a specific hypothesis. For example, Wang et al. \cite{Wang2014} studied the behavior of students in Dartmouth College using an ASys. During the semester, there were no app based or physical interference with students, and the application was primarily used for data collection, and their objective was to find out the relationship between behavior, emotions, and GPA during the studies. Similar traits can be seen in \cite{Santani2018, Biel2018, Moturu2011, Sano2013} too. Drawing motivation from self-monitoring related theories, in order to test the hypothesis in an unbiased manner, the participants of the smartphone sensing study should not have been aware of the hypothesis the researchers are testing for and they should not be allowed to change their behavior based on self-monitoring. Sagbas et al. \cite{Sagbas2020} explicitly mention this in their study, pointing out that they specifically did not let the participants know that they are conducting the study focusing on stress. As another example, if participants in \cite{Wang2014} knew that the experiment would be used to find connections with their GPA, there is a tendency that they would intentionally alter their behavior to obtain higher GPAs, hence having the effect of an interference. In \cite{Santani2018}, if students knew that researchers are testing for exactly alcohol drinking behavior, they might feel as if they are drinking less or more alcohol, hence alter the drinking behavior. Further, unlike research methodologies such as action research \cite{Lewin1946} where researchers engage with test participants to discuss and evaluate research techniques, analytical systems are better off being tested without self-monitoring capabilities (no feedback, interactions or interventions with participants), hence making sure that participants are sensed in their natural environments without any significant behavioral change due the usage of the mobile application.

While it is understood that the ultimate goal of smartphone sensing studies targeting the well-being of young adults is to infer well-being conditions just using passive sensing techniques, in order to get there, self-reports are still required. Hence, it involves smartphone users explicitly reporting conditions such as their emotions, stress levels, drink consumption, food consumption etc. When reporting such data for several days, it might bias the users leading to changes in the normal behavior of users. E.g. a person who reports drinking behavior might have reported that they did not drink alcohol in 2 weeks. When they realize this because they focus on it on a daily basis, it might motivate them to drink alcohol. Therefore, the real challenge in conducting smartphone sensing studies with analytical systems is that it is very difficult to reduce effects of self-monitoring during the study period. Current body of research does not discuss regarding this aspects in detail, and for smartphone sensing studies regarding well-being to grow in maturity for real world utility, aspects regarding self-monitoring should be thoroughly taken into consideration, and necessary principles should be adopted during participant recruitment phase and deployment phase to make sure that the biases are reduced and interventions are minimal during the study period. This would require novel recruitment policies for smartphone sensing studies that deal with young adults, special attention to reduce self-monitoring by carefully preparing the smartphone application, and its' content (e.g.: not revealing/emphasizing the final goals of a research, not showing statistics that might bias the behavior) and many more innovative techniques by considering the mindset of young adults and their behavior. 

\subsection{Feedback Systems}\label{subsubsec:feedbacksystems}
All the studies considered under this section provided feedback to users, or had behavioral intervention strategies included in the study. In implementation, studies used some sort of data analysis and visual representations that are visible to the users, making the system more of a real-time tool rather than a tool that is just used for data collection. Hence, these systems have allowed self-monitoring to users, making sure that the system is monitoring the progress or the reaction of the users when reflecting upon the feedback they get through the application. Hence, the goal of this kind of systems would be to evaluate the effectiveness of feedback, and the engagement and behavioral change from the perspective of users. It should be understood that when providing users with feedback, they reflect upon it, and alter their behavior more often than not. That is why the system perspective is important to be considered when designing smartphone sensing systems, specially for young adults. 

LiKamWa et al. \cite{LiKamWa2013} carried out a study regarding moods of young adults. Their study included components such as sharing user mood with others, viewing mood history, and they used a star achievement system to motivate users to self report more often, allowing them to collect a high amount of accurate ground truth data. It should be noted that even though this is not an intervention to the participants, this self-monitoring capability might cause behavioral changes in participants. Ma et al. \cite{Ma2012} too provided users with real-time feedback summarizing some information such as distance traveled, SMS frequency, most lasting activity, and general mood. Again, this feedback in the app is not necessarily an intervention, but allows users to self-monitor their behavior. Lin et al. \cite{Lin2012}, in their study \emph{BeWell+}, deployed a comprehensive real-time system, which involved in-device activity inference while community adapted well-being scores were computed in servers to provide user feedback. Further, they used a multi-dimensional ambient display with a fish to exhibit how well users are doing in-terms of the well-being score. They mentioned that their system received good feedback from 70\% of users at the end of the study. Their off-the-shelf analysis provided further insights saying that users of the system improved their well-being scores while using the app (positive behavioral change), elucidating the benefit of allowing the users to self-monitor. This is an example of how self-monitoring is activated in order to evaluate the usefulness of an end-to-end feedback system that facilitate behavioral change. 

Rabbi et al. \cite{Rabbi2015} have done a study regarding the well-being of a group of people including 10 young adults. They analyzed food, exercise, and activity patterns of users to provide user adapted well-being suggestions. They also analyzed how smartphone users make use of these feedback to enhance their well-being. Rodriguez et al. \cite{Rodriguez2017}, in their large scale study regarding smartphone user emotions, provided a simple feedback regarding emotional states of the users. Further, they introduced gamified elements to their feedback where app users were able to receive additional feedback by providing more self report data. This study is a proof that giving comprehensive feedback to obtain more user data in real world, in-the-wild deployments can be successful. Farhan et al. \cite{Farhan2016} have used real-time self reported data from their smartphone app to make clinical interventions, where they make sure that university students who have higher stress levels are interviewed by clinicians for them to continue to be part of the study.

Smartphone sensing applications are supposed to run unobtrusively as background apps, and it is important that these applications are optimized in-terms of performance for a better user experience. Fewer studies in our scope have done a thorough system analysis with the deployed smartphone application. LiKamWa et al. \cite{LiKamWa2013}, in their study regarding moods of young adults, have done an extensive analysis regarding their app "MoodScope" in smartphones such as Apple iphone 4 and Samsung Galaxy S2. For each phone, they computed latencies, computation times, power consumption, data consumption associated with background logging of data, in-phone data pre-processing and inference, and communication with servers. The only other study that has done a performance analysis of system aspects was by Rachuri et al. \cite{Rachuri2010}. In this study regarding emotion recognition with smartphone sensing, they have considered latency and energy consumption of their system with regard to speaker recognition and emotion recognition tasks. Even though certain studies \cite{Lin2012,Ma2012,Bae2017,Rodriguez2017} have discussed some performance aspects with regard to accelerometer sensing, location sensing, and battery life, they have not performed extensive analyses regarding these system aspects. As a summary, the user experience is an important component in feedback systems, because social science research too suggest that young adults have a taste towards interactive, gamified, less buggy, and feedback providing smartphone applications \cite{Park2018, Milward2016}.

\section{Discussion}\label{sec:discussion}

\subsection{Diversity-Aware research in ubiquitous and mobile computing}\label{subsec:diversity} Diversity in Machine Learning \cite{Gong2019v2} is an important topic that has grown in popularity during the last few years. While traditional machine learning focuses on data, model, and inference; diversity-aware machine learning has components such as data diversification (e.g. age, sex, country, culture, race, etc.), model diversification, and inference diversification. Specially in computer vision domain, these issues have been discussed in depth where some examples are biases in face datasets \cite{Vincent2018} and facial recognition systems \cite{Raji2019}, and studies finding that self-driving cars are more likely to hit Black-American people \cite{Brandom2018,Cuthbertson2019}. Considering studies within our scope, we discussed some diversity aspects in Section~\ref{study_summary}. We now identify a set of key limitations and areas that require future work. 

\subsubsection{Lack of diversity in a sample population} Using a population of diverse groups where each group is represented by few individuals is problematic. This might lead to models that poorly learn and lead to wrong conclusions. For example, in \cite{Lin2012}, a study to understand well-being had a small population of 27 people with 9\% from CS department, 34\% doctors/medical researchers, and 57\% graduate students. If we consider these sample cohorts, the behavioral routines of undergraduates, doctors, medical researchers and graduate students could be different. Hence, the resulting analysis about well-being might differ for different groups. The question would be whether the models could have actually considered these diversity aspects of people. Another issues is when the sample sets are not diverse enough. For example, stress levels of students could be different based on the  classes they take, or many other factors. In \cite{Huang2016}, the authors recruited all the students participating in the study from a psychology class, while the conclusions are made as if they were valid to all university students. Whether this is a fair conclusion remains a question.

\subsubsection{Gender-Biased training data} Some studies contain rather gender-biased sample populations. As pointed out by Santani et al. \cite{Santani2018} and also by a body of specialized literature, diversity aspects such as gender of young adults could play a role in determining their drinking patterns. Yet, some research has used heavily biased gender ratios. For example, \cite{Sano2013} had 18 young people out of which 15 were men. \cite{Wang2014} used a dataset of 48 people out of which 38 were men. From this perspective, and as gender biases in machine learning systems become more evident \cite{Scheuerman2019}, having a more in-depth analysis regarding the results with gender diversity in mind would be very informative and conclusive. A body of literature, both classic \cite{Brown1993} and recent \cite{Meegahapola2020a, Cirillo2020}, could guide how researchers can pursue directions related to gender. Mobile sensing research, especially in domains of mental health should concretely look into these gender biases.

\subsubsection{Geographical Diversity} Smartphone usage behavior of young adults across  countries can differ \cite{Mathur2017, Meegahapola2020a, Meegahapola2020b}, ranging from the type of apps they use to the time of phone consumption due to plethora of reasons such as cost of phones, unique lifestyles, and culture. For example, for young adults in western countries, Friday night would be a relaxing day where they drink and party, the situation could be totally different in Asian countries such as India where drinking is not socially accepted. This kind of geographical diversity has not been thoroughly studied in the literature so far, and it could be mainly because most studies were done in one or two countries with limited young adult participation. As EmotionSense \cite{Rachuri2010} and Sea Hero Quest \cite{SeaHeroQuest2019} have demonstrated, with wide smartphone and internet adoption \cite{Mallawaarachchi2020}, changes in the attitudes of people regarding using smartphones, and availability of app-based ecosystems, now it has become possible to conduct large-scale smartphone sensing for wider audiences. This also makes it possible to conduct smartphone sensing studies for young adults, specifically considering geographical diversity. 

\subsubsection{Personalization vs. Diversity-Awareness} Another important aspect regarding diversity-awareness is its distinction from personalization. While personalization focuses on building different models for different individuals from personal data, diversity-aware models consider diversity-aware features in building a single machine learning model \cite{Gong2019v2}. While personal models allow learning details regarding a single user, diversity-aware machine learning models allow learning hidden patterns/routines regarding diverse user groups. Moreover, the goal of diversity-aware learning is preventing or at least mitigating racial, age, gender, and other biases in artificial systems while also leveraging this diversity to build better models that benefit communities in different parts of the world.

Unlike computer vision datasets, smartphone sensing datasets in academic research so far have been comparatively smaller. Hence, applying diversity-aware machine learning techniques to these datasets can be challenging, often calling for specific techniques for oversampling or data augmentation \cite{Chawla2002}. However, as smartphone usage continues to grow among young adults, we believe that there is an opportunity to conduct studies, across many geographical regions, with diverse young populations. Hence, leveraging these technological developments to design smartphone sensing studies that integrate human diversity is timely. Developing new diversity-aware learning and inference techniques for heterogeneous smartphone sensing data could be the key to applications and systems that are both innovative and relevant to society. In summary, two RQs from the above analysis are:

\begin{itemize}[wide, labelwidth=!, labelindent=0pt]
\item[\textbf{RQ1:}] How to design mobile health sensing studies with human diversity as a key value, across geographic regions and population groups? What are the techniques available to measure diversity in datasets using machine-mediated models?
\item[\textbf{RQ2:}] How to build diversity-aware machine learning models for mobile health sensing that take into account the diversity aspects of people to provide them utility, but also to not discriminate based on diversity aspects? 
\end{itemize}{}

\subsection{Robust machine learning models against sensor failure in feedback systems.}\label{subsec:sensor_failure}
Many studies examined in this review mentioned that there are instances of incomplete data because of sensor failures and lack of self-reports \cite{Santani2018, Biel2018}. Yet, the machine learning architectures in the studies reviewed here do not attempt to examine inferences under sensor failure situations. This is a very important aspect relevant to mobile sensing studies in the open world. Further, in ASys, this situation will not cause issues because data cleaning and processing is done later to avoid data segments where some data sources are missing. However, for FSys, having a single model that uses all the features generated from the study to perform an inference can be troublesome in cases of sensor failure where analytics, interventions, and insights must be provided in real-time. Moreover, given that different types of features (e.g. app related features, accelerometer features, location features) have proven to perform the same inference with similar accuracies \cite{Santani2018, Biel2018}, it would be interesting to train different models using different feature sets, and build ensemble architectures that would select the best set of models to perform an inference, given the availability of data. Extending this line of work, another direction would be the use of different ML models to process different  feature types in real-time to perform the same inference (e.g. LSTMs for accelerometer data, random forest classifiers for tabular categorical data from self reports, convolutional neural networks to process images, etc.) and combining them as an ensemble. In summary, two RQs from the analysis we just presented are:

\begin{itemize}[wide, labelwidth=!, labelindent=0pt]
\item[\textbf{RQ3:}] How to build ensemble machine learning architectures to make feedback systems more robust against sensor failure?
\item[\textbf{RQ4:}] How to use different types of ML models in an ensemble architecture, and what are the benefits of this approach in feedback systems?
\end{itemize}{}

\subsection{Interaction sensing to better understand young adults}\label{subsec:interactive_sensing}

According to our analysis, there are three primary advantages of using interaction sensing over continuous sensing. They are: (1) \textbf{Sensor Failure and Calibration.} As mentioned in previous sections, continuous sensing techniques primarily rely on embedded sensors on the smartphone. Some of these sensors might lose their accuracy over time unless properly calibrated (e.g. location, accelerometer, gyroscope, etc.) \cite{Calibrate2019,Wenxuan2017} and some sensors fail over certain situations (e.g. when the phone is in airplane mode, when location is turned off, when bluetooth or Wifi is turned off, when microphone access is not given, etc.). However, such pitfalls are less frequent in interaction sensing (compared to continuous sensing), because interaction sensing requires user permission less often, and thereafter, such permissions are not turned on or off on a continuous basis, hence reducing the probability of failure; (2) \textbf{Heterogeneous Smartphones.} Continuous sensing techniques primarily process data from sensors whose sampling rates and accuracies might vary significantly based on phone type, operating system, quality of the sensor, and availability of certain specialized sensors \cite{Gong2019v3}. Hence, this heterogeneous nature poses challenges in the context of well-being applications that would be run on different phones in different contexts. However, interaction sensing focuses on using low-dimensional representations of data such as app usage, screen on/off, battery events, and typing events that are to some extent agnostic to the heterogeneous nature of smartphones, making it possible to build ML models that require fewer adjustments and calibrations to function under different circumstances; and (3) \textbf{Young Adults and Smartphone Usage Behavior.} The current work in the domain reflects that priority has been given to continuous sensing techniques as summarized in Table~\ref{tab:sensing}. It should be understood that these sensing techniques act in a similar way regardless of the user, as in whether they are young adults or older people, and hence, the sensing data can be obtained with good quality regardless of the age group, as long as the technological infra-structure supports it (e.g.: wifi access, quality of smartphone, etc.). {However, when we consider interaction sensing, a key issue is how people use and interact with the smartphone.} Prior research \cite{PewResearchCentre2019, BankMyCell2019} suggests that there are significant differences in smartphone and app usage behavior of young adults compared to older generations. Hence, this might lead to signals of different strengths when used on young adults as compared to older generations. We believe that researchers in ubiquitous computing and human-computer interaction should focus on this aspect. Leveraging this prominent characteristic would allow a paradigm shift from continuous sensing to interaction sensing. While we understand the importance of continuous sensing techniques, interaction sensing offers a lot of room for improvement as mentioned above, and a multitude of open research questions in smartphone sensing, especially when targeting young adults and their well-being. In summary, two RQs from the previous discussion are:

\begin{itemize}[wide, labelwidth=!, labelindent=0pt]
\item[\textbf{RQ5:}] How to tackle methodological pitfalls of continuous sensing (e.g. sensor failure, heterogeneous sensors) by using different interaction sensing approaches focused on  health and well-being conditions?
\item[\textbf{RQ6:}] How to improve interaction sensing techniques to better suit different age groups?
\end{itemize}{}

\subsection{Making better use of knowledge in human sciences}\label{subsec:assessment_reactivity}
As per prior research we discussed in this study, it is unclear whether computer science researchers are fully making use of principles and techniques from human psychology, clinical, and behavioral research into smartphone sensing research. Given that smartphone sensing systems capture the behavior of young adults, it is essential to understand behavioral dynamics and psychology of this age group. When looking at this problem from the system perspective, two points arise: 
(1) \textbf{ASys:} If the requirement is to understand behavior or person aspects in everyday life, if researchers let app users know the test hypothesis explicitly (e.g. that we are testing for stress, happiness, heavy alcohol usage) and allow self-monitoring, it might bias the mindset of people (e.g. I reporting that I am stressed everyday, or that I should relax today). These biases may get reflected in the data if people alter their behavior, resulting in researcher's drawing possibly wrong conclusions from the studies.
(2) \textbf{FSys:}  If we want to understand how the behavior of users changes over time due to the usage of smartphone sensing-based feedback systems, it is necessary to influence users by using self-monitoring mechanisms using interventions and feedback. A clear example for this is BeWell system \cite{Lin2012}, which measure how well young people adapt their lifestyle based on the feedback they get. Another example is the work of Farhan et al. \cite{Farhan2016}, which generated real-time clinical interventions for participants who reported higher stress levels. 

These two types of systems are the two corners of the spectrum, and there can be studies which have features of both. We believe that proper attention should be given to system design and experimentation by drawing principles from behavioral research, keeping in mind what each type of system offers in terms of testing hypotheses. Researchers should be aware of how experiments get affected by human bias, behavior, and psychology.

It should also be highlighted that FSys are better used when the basic hypothesis regarding human behavior has already been established using ASys or clinical methodologies. For example, if the experiments are done with a FSys to analyze the relationship between well-being and physical activity levels, the setting might lead to wrong conclusions because self-monitoring capability of the applications alter the behavior of young adults. This kind of relationships are better to be examined first with ASys to establish relationships, and then test for behavioral change that is occurring using FSys.  
A summary of some research questions that can be derived from this section are:

\begin{itemize}[wide, labelwidth=!, labelindent=0pt]
\item[\textbf{RQ7:}] How to systematically decide when ASys or FSys are the best methodological choice for a given problem?
\item[\textbf{RQ8:}] How to leverage knowledge in human sciences to quantify, control, and reduce possible influences on behavior due to the use of ASys?
\end{itemize}{}

\subsection{Underused pillars of data and sensor types}\label{subsec:underused}
Table~\ref{tab:sensing} examines which sensor (passive sensing modality) is used as a proxy for a variable belonging to each pillar of data, and Table~\ref{tab:reports} examines which self-report modality has been used to get data. Both tables provide interesting perspectives to understand research gaps in the mobile sensing applications that target young adults. Table~\ref{tab:sensing} shows that accelerometer has been directly used as a proxy to variables belonging to the behavior pillar. The table also shows an opportunity as accelerometer has not been used as a direct proxy to the person pillar. Moreover, it also shows that interaction sensing modalities have rarely been used as proxies for the person or context pillars, that opens up opportunities for mobile health research to leverage such relationships (established in clinical research), and to use such findings in applications that target well-being aspects of young adults. Further, Table~\ref{tab:reports} also opens up interesting avenues for future mobile sensing research. As examples, pictures and videos, as well as audio have been used less frequently as proxies for variables belonging to the person or context pillars. This provide opportunities to assess whether such relationships exist in other research domains (e.g. inferring ambiance of a surrounding place using a video/photo, etc), and use such relationships in mobile sensing studies. As a summary, two RQs for future work are: 

\begin{itemize}[wide, labelwidth=!, labelindent=0pt]
\item[\textbf{RQ9:}] Can continuous sensing techniques (e.g. accelerometer, proximity, gyroscope, location, ambient light, audio, bluetooth, wifi, etc.) be used as proxies to attributes that belong to the person pillar (e.g. stress, emotions, depression, sociability, mood, happiness, etc.)?
\item[\textbf{RQ10:}] Can interaction sensing techniques (e.g. app usage behavior, typing events, touch events, battery levels, phone calls, messages, screen events, etc.) be used as proxies to attributes that belong to the context pillar (social context, semantic location, ambient light, temperature, crowdedness, etc.)?
\end{itemize}{}

\subsection{participant recruitment strategies}\label{subsec:recruitment}
We have highlighted several participant recruitment strategies used in this set of studies in Section~\ref{study_summary}. As mentioned in that section, it is evident that computer science researchers have not paid enough systematic attention to participant recruitment strategies. Most studies used convenience samples without a more formal recruitment strategy. We believe that one open area of research would be to systematically compare different recruitment strategies from the perspective of utility for mobile sensing-based research. Research along this line would not only help research targeting young adults, but also mobile sensing research as a whole. While smartphone sensing studies in our scope have focused mainly on {cost} in terms of money and time when recruiting participants, human science research focuses on aspects such as scientific validity and ties with related clinical research \cite{Berger2009,Patel2003}. Moreover, in social science research, in addition to proper recruitment strategies, corrections are used to unbalanced variables such as age, gender, or ethnicity with rigorous statistical analysis \cite{Elhai2017}. This makes sure that results are scientifically valid.

\begin{itemize}[wide, labelwidth=!, labelindent=0pt]
\item[\textbf{RQ11:}] Do datasets collected from mobile health sensing studies from young adults reflect different recruitment strategies? What are the implications on data when not using systematic recruitment strategies as compared to clinical studies?
\end{itemize}{}

\subsection{Using contextual triggers to collect self-reports}\label{subsec:whichwhere} 
When understanding the notions of each of these strategies from the perspective of pillars of data, some interesting observations can be made. Aspects that are often subjective, such as emotions, feelings, and stress, which belong to the person pillar, are difficult to be captured retrospectively. For example, asking a young adult how did they feel when they were in the lecture hall might introduce recall bias because socio-psychological notions are often subjective. On the other hand, self-reporting aspects such as eating or drinking behaviors that lead to objective answers might be easier to collect retrospectively, and in some cases they might be cumbersome to collect in-situ. For example, if young adults are asked to explicitly report their drinking behavior (data belonging to behavior and context pillars), in-situ reporting would be cumbersome to them if they are dancing in the club and it might lead to low response volumes. In fact, Bae et al. \cite{Bae2017}, have experimentally reported that retrospective reporting of alcohol drinking behavior have provided better results compared to in-situ reporting. In general, it can be said that, the more intimate the self-reported activity is, the more thought should be given to the subjectivity of such data, the privacy of users, and the ease of use of the application. While subjective data are better off being collected in-situ to avoid bias, the trade-off is that the application might be adding a layer of complexity. Another consideration regards the actual motivations of young adults in using the application. If the users are motivated to achieve a personal goal (e.g.: lose weight, reduce alcohol usage, eat healthy food), such users might not need push notifications reminding them to report their details. On the other hand, if the study participants are not intrinsically motivated to report certain aspects, then reminders might be useful.

\begin{itemize}[wide, labelwidth=!, labelindent=0pt]
\item[\textbf{RQ12:}] How can contextual triggers be used to improve mobile sensing data collection? What are the effects of these methodological choices on the final outcomes of a study?
\end{itemize}{}

\subsection{Transfer Learning and Meta-Learning} 

In computer vision problems, transfer learning is commonly used with the expectation that convolutional neural networks would learn certain features that would help classify other related datasets \cite{Pan2010}. However, mobile health sensing data that are time series and tabular in nature are sparse and complex, and features of datasets are manually crafted, hence making transfer learning and dataset concatenation difficult \cite{Ma2020, Gong2019v3}. As an example, mobile health applications contain passive sensing information (high dimensional and high resolution) and self-reports (low dimensional). Moreover, different research studies analyzing similar aspects (e.g. food consumption \cite{Seto2016, Biel2018}, stress \cite{Lu2012, Sano2013, Bogomolov2013}, depression \cite{Boukhechba2018, Farhan2016, Canzian2015}, and drinking episodes \cite{Bae2017, Santani2018}) have collected these behavioral traces using different protocols, phone types, and techniques resulting in datasets with contrasting characteristics and different types and number of features. However, the fundamental types of data that are collected in behavioral smartphone sensing are under three main pillars of data P, B, and C according to the taxonomy we proposed. The general expectation is that these pillars would show high correlation across datasets that are similar in nature and analyze the same behavioral aspect. As an example, two studies regarding \emph{stress} could use raw accelerometer traces or steps counts (derived in the smartphone using IMU sensors) belonging to B pillar to indicate the physical activity levels. Even though the collected data are different in the number and type of features, what they capture is the physical activity level, and we could expect those features to have similar correlations with stress. Hence, one approach could be to craft intermediate data representations from features belonging to similar pillars to seek the possibility of transfer learning, hence accommodating transfer learning among datasets in the same pillar and domain, while the collected datasets come from using slightly different protocols. 

In addition to the challenging use-case mentioned above, there could be instances where training and testing distributions could differ, even if the same set of features are collected using the same procedure in training and deployments phases \cite{Hernandex2020, Ma2020, Gong2019v3}. An example is a mobile sensing inference pipeline to count steps in android phones developed using data collected in the United States, being deployed in China. One could expect differences in the type of phones that are used (e.g. Samsung, Huawei, Sony, Xiomi, etc.), and also the way people use the phone, hence leading to different and inaccurate step count results when deployed in the wild, even if the same set of features are collected. In a recent study, Gong et al. emphasized this issue and provided a solution to the problem using meta-learning \cite{Gong2019v3}, showing that their technique MetaSense achieves state-of-the-art results even compared to other meta learning and transfer learning techniques for activity recognition and speech recognition tasks. Studies along this line would help this research in making sure that datasets and models in the domain are re-usable, hence accelerating the growth of the research field. While most of the recent work \cite{Gong2019v3, Ma2020} are focused on speech and accelerometer sensors, it would be worth to look at other commonly used modalities in mobile health sensing. Moreover, this could be another way of accommodating diversity-awareness (discussed in Section~\ref{subsec:diversity}) in smartphone sensing research through machine learning.

\begin{itemize}[wide, labelwidth=!, labelindent=0pt]
\item[\textbf{RQ13:}] How to build intermediate representations from mobile sensing data to allow transfer learning to different tasks, even when datasets are collected using different protocols? 
\item[\textbf{RQ14:}] How can meta-learning be used for heterogeneous data collected in mobile sensing applications, when datasets are collected using the same protocol but from diverse population groups? 
\end{itemize}{}

\subsection{Other Issues}\label{subsec:otheryouthissues}

In this study, we came across literature in pillars P and B, for domains such as stress, depression, anxiety, alcohol consumption, eating behavior, etc. as shown in Table~\ref{tab:pillars_domains}. In addition, there exists opportunities for researchers to expand upon the current work that are focused on P and B pillars, and to examine whether smartphone sensors could be used to characterize contextual attributes that are linked to domains in P and B. Consider the following example regarding eating behavior for clarity. Smartphone sensing involving eating behavior takes the form of mobile applications that combine passive smartphone sensing with mobile food diaries: (1) Detecting eating events - this type of studies could be done with the aim of detecting the time of eating. Hence, a smartphone should be able to understand the time at which a user would eat. This task has been previously attempted using wearable sensing \cite{Rahman2016, Edison2015}. Hence, given the time of the day, the smartphone should be able to determine whether a user is eating or not; (2) Characterizing eating events - food types and categories are the most important aspects regarding the food consumption. A smartphone application that could unobtrusively infer categories of food taken by users depending on sensed contextual cues could be a powerful tool in providing users with valuable interventions and analytics. The study by Biel et al \cite{Biel2018} belongs to this category. In addition, this step could go beyond the consumed food type, and attempt to characterize the eating event in more depth by understanding food consumption levels and the contextual aspects that affect food consumption. For example, aspects such as the social context of eating (an contextual attribute belonging to pillar C) is known to affect the food consumption behavior of people according to concepts such as social facilitation and impression management \cite{Herman2003, Herman2005, Meegahapola2020b}. The social context also affects the amount of food people eat -- i.e. highly social contexts can lead to overeating \cite{Hetherington2006, Meegahapola2020b}. If smartphone sensing capabilities could be used to sense the social context of eating unobtrusively, such inferences could be used to provide users with context-aware interventions -- e.g. if a highly social eating context is sensed, notify the users to be aware of the food intake.

As shown by the two steps above, even for other domains mentioned in Table~\ref{tab:pillars_domains}, a similar framework could be used to explore research directions for future work. In general terms, the two main steps that we identified are: \textbf{(1) Event Detection} - given the \emph{time} of the day, smartphone should be able to unobtrusively infer whether an event is occurring. Some examples for events are feeling happy, feeling stressed, not feeling depressed, eating food, and drinking alcohol; once the events are detected, the next step is to \textbf{(2) Event Characterization} - given that an event is detected (the time of the event is known), smartphones could infer attributes related to the context of the event. For a domain like depression, the characterizations could be related to the ambiance of the environment or the social context that might affect a person. For a domain such as alcohol consumption, characterizations could be related to social context of drinking or types of drinks that are consumed (e.g. beer, wine, other spirits, etc.). Similarly, understanding smartphone sensing research done in various domains using the above mentioned two-fold categorization would allow future research to go into more depth in each of the domains. Hence, according to findings in Table~\ref{tab:pillars_domains}, smartphone sensing research could look into characterizing events in domains where detection has been shown to be feasible.

There are several other well-being related areas that are common among young adults, but discussed relatively less in smartphone sensing research according to Table~\ref{tab:pillars_domains}. For example, aspects such as anger and violence, have been identified as important aspects that affect young adults according to prior literature \cite{Anger2019}, but have not been given proper attention in smartphone sensing research. In addition, there are certain domains that have been discussed in human-computer interaction (HCI) literature, but have not been studied in smartphone sensing. For example, gender specific issues such as menstruation, while being discussed in HCI literature for sometime \cite{Epstein2017}, has not been studied in smartphone sensing research. Moreover, the study by Epstien et al. \cite{Epstein2017} emphasized that currently available menstrual cycle tracking apps (that do not use passive sensing) do not take into account diversity aspects that affect the menstrual cycle length such as young adulthood, pregnancy, and menopause. This again is an example of proof of why smartphone sensing techniques should be designed with diversity-awareness in mind. Furthermore, self-harm is another well-being aspect that has been discussed in HCI literature as a common issue among young people \cite{Honary2020}, but has not been given enough emphasis in smartphone sensing studies. Finally, some other issues faced by young adults that have not been addressed using smartphone sensing research include self-esteem \cite{SelfEsteem2019}, obsessive compulsive disorder \cite{Weissman1998}, and addictive behaviors \cite{Porn2019, Lauvsnes2020}, for which there are both a body of literature and active work in public health research.

\begin{itemize}[wide, labelwidth=!, labelindent=0pt]
\item[\textbf{RQ15:}] Even though mood instability and stress have been discussed in smartphone sensing studies regarding young adults, aspects such as self-harm, anger, and violence have not been given proper attention in the domain. Can smartphone sensing and machine learning models, integrated with ethical and privacy-by-design principles, be used to support young adult populations regarding these situations? 
\end{itemize}{}

\subsection{Ethical Issues on Future Smartphone Sensing}\label{subsec:privacy}
As the discussion above highlights, the use of smartphone sensing for studies involving young adults raises a number of ethical issues. First, personal data is involved in the studies reviewed here, including both sensor data like location, as well as survey data where participants disclose information including their habits, state of mind, and even sensitive information such as age, gender, etc. \cite{Meegahapola2020a}. Furthermore, some of the reviewed studies address rather sensitive issues like mental health. Clearly, this research requires both legal and ethical review procedures, which are well established in many countries, although not in the whole world. The existence of recent regulations like the European General Data Protection Regulation (GDPR), and similar efforts in other parts of the world, provide frameworks to address some of these issues, but also uncover potential inconsistencies across world regions. This is especially important for studies that aim to engage diverse populations across countries. Second, many of the current concerns about machine learning-based decision making systems are also applicable to the mobile sensing setting, including issues of fairness, accountability, and transparency. This is an issue that will become increasingly important. Finally, the commercial use of smartphone sensing apps to monitor aspects of young adults well-being also poses ethical questions, where economic interests of companies may not always align with the objective of supporting young adults to achieve and sustain their own goals regarding well-being (see for instance recent journalistic coverage of mobile health apps for college students in the US \cite{Siegel2019,Paul2019, Harwell2019}). These fundamental issues need to be considered by researchers as future studies and systems are designed and deployed.

\section{Conclusion}

In this paper, we extensively analyzed studies that leverage smartphone sensing for the well-being of young adults. By specifically targeting young people, we narrowed down the type of well-being related aspects, and hence identified how self-reports and passive sensing of smartphones have been applied in a very specific context. First, we used human-science concepts and theories to frame behavioral research in smartphone sensing using a solid framework. We analyzed the Data Perspective of smartphone sensing using pillars of data, and the System Perspective using self-monitoring related concepts. We emphasized aspects including, but not limited to the importance of focusing on interaction sensing to better understand young adults; feedback systems to test for behavioral change; analytical systems to test for hypothesis regarding specific behavioral aspects; diversity-awareness of smartphone sensing to support diverse groups of people; developing feedback systems that are robust to sensor failure; and using systematic recruitment strategies for mobile sensing studies regarding young adults. 

\bibliography{bibtex/bib/IEEEabrv.bib,bibtex/bib/IEEEexample.bib}{}
\bibliographystyle{IEEEtran}

\begin{IEEEbiography}[{\includegraphics[width=1in,height=1.25in,clip,keepaspectratio]{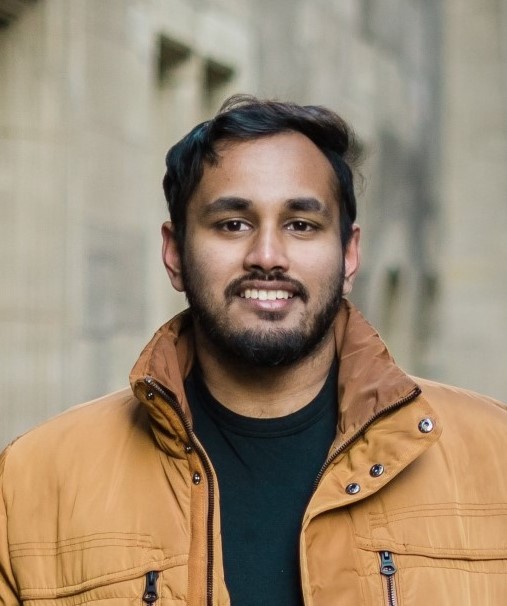}}]{Lakmal Meegahapola} was born in Kandy, Sri Lanka in 1993. He received his B.Sc. in Computer Science and Engineering from the Faculty of Engineering, University of Moratuwa, Sri Lanka in 2018. Currently, he is a Doctoral Candidate specializing in Mobile Sensing at the School of Engineering, École polytechnique fédérale de Lausanne (EPFL), Switzerland and a Research Assistant in the Social Computing Group at Idiap Research Institute, Switzerland. He is interested and experienced in using machine learning and data mining techniques for research in the intersection of (a) mobile health sensing, (b) context-awareness, and (c) social computing.
\end{IEEEbiography}

\begin{IEEEbiography}[{\includegraphics[width=1in,height=1.25in,clip,keepaspectratio]{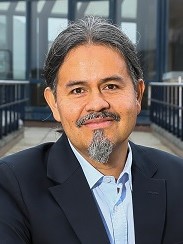}}]{Daniel Gatica-Perez} directs the Social Computing Group at Idiap Research Institute, and is also a Professor at EPFL, Switzerland. His research interests span social computing, ubiquitous computing, and crowdsourcing for social good. He is a member of the IEEE.

\end{IEEEbiography}

\EOD

\end{document}